\documentclass[doublespacing]{elsart}
\usepackage[usenames]{color}
\usepackage{amsmath,amssymb,amsfonts}
\usepackage{latexsym}
\usepackage{textcomp}
\usepackage{pifont}
\usepackage{dcolumn}
\usepackage{hhline}
\usepackage{rotating}
\usepackage{multirow}
\usepackage{tipa}
\usepackage{ae,aecompl}
\usepackage{mathrsfs}
\usepackage{colortbl}
\usepackage{longtable}
\usepackage{lineno}
\usepackage{framed}
\usepackage{natbib}
\begin{document}
\begin{frontmatter}
%%%%%%%%%%%%%%%%%%%%%%%%%%%%%%%%%%%%%%%%%%%%%%%%%%%%%%%%%%%%%%%%%
% Title Page.                           
%%%%%%%%%%%%%%%%%%%%%%%%%%%%%%%%%%%%%%%%%%%%%%%%%%%%%%%%%%%%%%%%%
\title{The opposition effect in the outer Solar System: a comparative study of the phase function morphology}
%\vspace*{6ex}
%\author{Estelle \ \textsc{D\'eau}\footnotemark[1]~\textsuperscript{a}, Luke  \ \textsc{Dones}\textsuperscript{b}, S\'ebastien \ \textsc{Charnoz}\textsuperscript{a}, Andr\'e \ \textsc{Brahic}\textsuperscript{a}, and Carolyn C. \ \textsc{Porco}\textsuperscript{c}\vspace*{6ex}}
%\affil{\textsuperscript{a} AIM, UMR 7158  Service d'Astrophysique CEA Saclay 91191 Gif-sur-Yvette FRANCE} 
%\affil{\textsuperscript{b} Southwest Research Institute, 1050 Walnut Street, Suite 300 Boulder CO 80302, USA} 
%\affil{\textsuperscript{c} CICLOPS, 3100 Marine Street Suite A353, Boulder CO 80303, USA \vspace*{6ex}}
\author[deau]{Estelle D\'eau\textsuperscript{1,}}, 
%\author[deau]{Estelle D\'eau\footnote[*]{Direct editorial correspondence to~: Estelle \textsc{D\'eau}, e-mail : edeau@cea.fr, Phone~: +33 (0) 1 69 08 80 56, Fax~: +33 (0) 1 69 08 65 77}},
\author[dones]{Luke Dones}, %,
\author[deau]{S\'ebastien Rodriguez}, 
\author[deau]{S\'ebastien Charnoz} and
\author[deau]{Andr\'e Brahic}
%\author[porco]{Carolyn C. Porco}
\address[deau]{CEA/IRFU/Service d'Astrophysique, AIM UMR 7158, 91191 Gif-sur-Yvette FRANCE}
\address[dones]{Southwest Research Institute, 1050 Walnut Street, Suite 300 Boulder CO 80302, USA}
%\address[rodriguez]{CEA/IRFU/Service d'Astrophysique, AIM UMR 7158, 91191 Gif-sur-Yvette FRANCE}
%\address[charnoz]{CEA/IRFU/Service d'Astrophysique, AIM UMR 7158, 91191 Gif-sur-Yvette FRANCE}
%\address[brahic]{CEA/IRFU/Service d'Astrophysique, AIM UMR 7158, 91191 Gif-sur-Yvette FRANCE}
%\address[porco]{CICLOPS, 3100 Marine Street Suite A353, Boulder CO 80303, USA}
\end{frontmatter}
\footnotetext[1]{Corresponding author~: Dr. Estelle \textsc{D\'eau}, \\
Address~: CEA/IRFU/Service d'Astrophysique, Orme des Merisiers Btiment 709, 91191 Gif-sur-Yvette FRANCE, \\
e-mail estelle.deau@cea.fr, Phone +33(0)1 69 08 80 56, Fax +33(0)1 69 08 65 77} 
 
%\maketitle
%\begin{center}
%\end{center}
%\vspace*{2ex}
%
\vspace{1cm}
Number of pages: \pageref{lastpage} \\
Number of tables: \ref{lasttable} \\ 
Number of figures: \ref{lastfig} \\
Running head: The opposition effect in the outer Solar System %\vspace*{2ex}
%\end{flushleft}
%\noindent keywords: Planetary rings -- phase curves; opposition effect -- coherent
%backscattering, shadowing, shadow-hiding %\vspace*{2ex}
%\newpage
%
%\noindent Direct editorial correspondence to~: \\
%Estelle \textsc{D\'eau} \\
%CEA/IRFU/Service d'Astrophysique  \\
%AIM UMR 7158  \\
%Orme des Merisiers \\
%91191 Gif-sur-Yvette FRANCE \\
%phone: +33 1 69 08 80 56 \\
%fax: +33 1 69 08 65 77 \\
%e-mail : edeau@cea.fr \\

%%%%%%%%%%%%%%%%%%%%%%%%%%%%%%%%%
%    Abstract
%%%%%%%%%%%%%%%%%%%%%%%%%%%%%%%%%
\newpage
\begin{linenumbers}
\begin{center}
{\bf ABSTRACT}
\end{center}
%\begin{abstract}
In this paper, we characterize the morphology of the disk-integrated phase functions of satellites and rings  
around the giant planets of our Solar System. We find that the shape of the phase 
function is accurately represented by a logarithmic model \citep[][in {\em Surfaces and Interiors of Planets and Satellites}, Academic, edited by A. Dollfus]{1970sips.conf..376B}. 
For practical purposes, we also parametrize the phase curves by a linear-exponential model \citep[][{\it Journal of Quantitative Spectroscopy and Radiative Transfer}, {\bf 70}, 529--543]{2001JQSRT..70..529K} and a simple linear-by-parts 
model \citep[][{\it Astronomical Journal}, {\bf 81}, 865--893]{1976AJ.....81..865L}, which provides three 
morphological parameters~: the amplitude~$A$ and the Half-Width at Half-Maximum (HWHM) of 
the opposition surge, and the slope $S$ of the linear part of the phase function at larger 
phase angles. \\
Our analysis demonstrates that all of these morphological parameters are correlated with 
the single scattering albedos of the surfaces. \\ 
By taking more accurately into consideration the finite angular size of the Sun, we find that the Galilean, Saturnian, Uranian and Neptunian 
satellites have similar HWHMs ($\lesssim$0.5\textsuperscript{o}), whereas they have a wide range of amplitudes~A. The Moon has the largest HWHM ($\thicksim$2\textsuperscript{o}). 
We interpret that as a consequence of the ``solar size bias'', via the finite size of the Sun which varies dramatically from the Earth to Neptune. By applying a new method that attempts to morphologically deconvolve the phase function to the solar angular size, we find that icy and young surfaces, with active resurfacing, have the smallest values of A and HWHM, whereas dark objects (and perhaps older surfaces) such as the Moon, Nereid and Saturn's C ring have the largest A and HWHM. \\
Comparison between multiple objects also shows that Solar System objects 
belonging to the same planet host have comparable opposition 
surges. This can be interpreted as a ``planetary environmental effect'' that acts to modify locally the regolith and the surface properties of objects which are in the same environment. 

%\end{abstract}
%\end{frontmatter}
\noindent \textit{keywords}: Planetary rings; Satellites of Jupiter, Saturn, 
Uranus, Neptune, phase curves; opposition 
effect, coherent backscattering, shadowing, shadow-hiding, angular size of the solar radius %\vspace*{2ex}
%\end{keyword}
%\newpage
%~~
%\newpage
%\vspace*{2ex}
%\tableofcontents

%%%%%%%%%%%%%%%%%%%%%%%%%%%%%
% Introduction
%%%%%%%%%%%%%%%%%%%%%%%%%%%%%
\newpage
\section{Introduction} \label{intro} 
The opposition effect is a nonlinear increase of brightness when the phase angle $\alpha$
(the angle between the source of light and the observer as seen from the body) decreases to zero. This 
effect was seen for the first time in Saturn's rings by \citet{1884AN....109..305S} 
and \citet{1885AN....110..225M}. Now, this photometric effect has been observed on  
many surfaces in the Solar System~: first on satellites of the giant planets, see 
\citep{1997Icar..128....2H} for a review; second on asteroids, 
\citep{1989Icar...77..171H,1989Icar...81..365H,2000Icar..147...94B} and Kuiper Belt Objects \citep{2008ssbn.book..115B,2002ocd.book..224}; and finally on various surfaces on Earth \citep{1990Icar...88..418V,1996LPI....27..491H} and for minerals in the laboratory 
\citep{1999Icar..141..132S,2003AA...409..765K}. 
The opposition effect on bodies in the Solar System 
has supplied interesting constraints about the regolith and state of the
surfaces \citep{1997Icar..128....2H,M2006}. Indeed, the opposition effect is now 
thought to be the combined effect of coherent backscatter (at very small phase angles), 
which is a constructive interference between grains with sizes near the wavelength of light, and 
shadow hiding (at larger phase angles), which involves shadows cast by the 
particles themselves \citep{1997Icar..128....2H}. \\
By parametrizing the morphology of the phase functions for $\alpha\thicksim$ 
0--20\textsuperscript{o}, some numerical models have derived physical properties 
of the medium in terms of regoliths \citep{1992MNRAS.254P..15M,1999Icar..141..132S} 
and the state of the macroscopic surface \citep{1986Icar...67..264H,2002Icar..157..523H,1999Icar..141..132S}.
However, such characterization of the phase function morphology is restricted by 
the angular resolution and the phase angle range of the observed phase function. 
Moreover, some effects (such as the finite size of the Sun and the nature of the soil), 
which are not yet taken rigorously into account by the most recent models, can play 
important roles in a comparative study. \\
For these reasons, it seemed important to test the behavior of the morphology 
of the phase function before using any physical model.

The use of a simple morphological model is generally not adapted to derive the physical 
properties of the medium. But for the data set presented here, only the disk-integrated brightness 
I/F and the phase angle~$\alpha$ are available, the corresponding angles of incidence ($i$) 
and angles of emission ($\epsilon$) are not given for these observations, so we cannot use 
sophisticated for further investigations analytical models 
\citep{1986Icar...67..264H,2002Icar..157..523H,1999Icar..141..132S} which need the 
brightness~$I/F$ and the three viewing geometry parameters~$\alpha$, $\mu$ and 
$\mu_{0}$ ($\mu$ and $\mu_{0}$ are the cosines of $\epsilon$ and $i$, respectively). \\
However, the theories developed for the coherent-backscattering 
and the shadow-hiding effects deduce their properties by parametrizing the opposition phase curve  \citep{1992MNRAS.254P..15M,1992Ap&SS.189..151M,1992Ap&SS.194..327M,1999Icar..141..132S,1986Icar...67..264H,2002Icar..157..523H}. 
Thus it is possible to connect the morphological parameters A, HWHM and S with some physical 
characteristics of the medium derived from these models.

The amplitude A of the opposition peak is generally known to express the effects of the coherent-backscattering. According to \citet{1999Icar..141..132S,2000Icar..147..545N}, A is a function of grain size in 
such way that A decreases with increasing grain size (we refer to grains as the smallest scale of the 
surface compared to the wavelenght and virtual entities implied in the coherent backscatter effect, as microscopic roughness). This anti-correlation finds 
a natural explanation in the fact that for a macroscopic surface, large irregularities 
with respect to the wavelength create less coherent effects than  irregularities with sizes comparable to the wavelength. \\  
\citet{1992Ap&SS.189..151M} and \citet{1992Ap&SS.194..327M} emphasize that 
A is linked to the intensity of the background $I_{b}$ \citep[defined as a morphological parameter of the linear-exponential function of][]{2001JQSRT..70..529K}, which is a decreasing 
function of increasing absorption \citep{1990AdSpR..10..187L}; thus A must 
increase with increasing absorption or decreasing albedo~$\varpi_{0}$, which 
was confirmed by the laboratory measurements of \citet{2003AA...409..765K}.  
Indeed, \citet{2003AA...409..765K} remarked that the opposition surge 
increases and sharpens when irregularities are small and that the opposition 
surge decreases with increasing sample albedo.    

The half width at half maximum HWHM, is also associated to the coherent-backscatter 
effect. It has been related to the grain size, index of refraction, and packing density of regolith, by previous numerical studies \citep{1992Ap&SS.194..327M,1992MNRAS.254P..15M,2002Icar..157..523H}. 
The variation of HWHM with these three physical parameters is complex; 
see Fig.~9 of \citep{1993ApJ...411..351M}: HWHM reached its maximum for an effective 
grain size near $\lambda$/2 and increases when the regolith grains' filling factor~$f$ 
increases. For high values of $f$, the maximum of HWHM occurs for a larger grain size. \\
However, several studies \citet{1997Icar..128....2H,2000Icar..147..545N,2002Icar..157..523H} defined two HWHM parameters~: for the \citet{2002Icar..157..523H} model, the 
coherent-backscatter HWHM ($h_c$), which is defined similarly to that in the model of \citet{1992Ap&SS.194..327M}, and the shadow hiding parameter $h_s$. Applying this model to Saturn's rings, 
\citet{2007PASP..119..623F} found that the coherent-backscatter peak is about ten times narrower than the shadow-hiding peak, but neither  $h_c$ nore $h_s$ equals to the morphological width of the peak HWHM. This reinforces the idea that a coupling of the two opposition effect mechanisms at small phase angles could be responsible for the observed surge width.    

Since the efficient regime of the shadow hiding is 10\textsuperscript{o}--40\textsuperscript{o}\citep{1985Icar...64..320B,1997Icar..128....2H,S1999} and that of the coeherent bakscattering does not exceed several degrees \citep{1997Icar..128....2H}, the slope of the linear part $S$ can be regarded as the only parameter that mirrors the shadow hiding solely. This slope depends on the particle filling factor~D, which relates to the porosity of the regolith of a satellite and the ratio between the particle size and the physical thickness of the ring for a planetary ring \citep{1966JGR....71.2931I,S1999,1974IAUS...65..441K}. 
For a satellite, by ``particles'' we mean the macroscopic scales of the surface, which are implied in the shadow hiding effect. \\
In the shadowing model of \citet{1966JGR....71.2931I} and \citet{1974IAUS...65..441K} 
(which consists of the effects of shadows for a monolayer of particles), when the slope 
is shallow, the  variation of $\alpha$ does not change the visibility of shadows and 
the particle filling factor must be high to make 
the proportion of shadows small for any observation geometry. By contrast, when S 
is steep, the particle filling factor is 
smaller and will contribute to a broad and large peak with a weak amplitude 
which will be regarded as a slope. \\
In the shadow hiding model (i.e., multilayer shadowing), the larger the optical depth and the volume 
density~D, the steeper the phase function is at large phase angles \citep[10\textsuperscript{o}--40\textsuperscript{o},][]{S1999}. However, at larger phase angles, the behavior of the absolute slope with albedo could change according to a more efficient regime of the shadow hiding (50\textsuperscript{o}--90\textsuperscript{o}, Stankevich~2008, \textit{private communication}).  

For a compact medium such as a satellite's surface, the slope at very large angles 
($\alpha$>90\textsuperscript{o}) is a consequence of topographic roughness, the so-called roughness 
parameter~$\theta$ in the \citet{1984Icar...59...41H,1986Icar...67..264H} model. 
Then a steeper slope is due to a surface tilt which varies from millimeter to centimeter 
scales \citep{1984Icar...59...41H}. However, the roughness can influence the phase curve at smaller phase 
angles as underlined by \citet{1985Icar...64..320B}, also according to the laboratory measurements of \citet{2003AA...409..765K}, 
the slope of the phase function ($\alpha$<40\textsuperscript{o}) increases with increasing roughness. 
From the theoretical assertions made above, the HWHM and the amplitude are governed 
by both coherent backscatter and shadow hiding effects, whereas the slope of the 
linear part of the phase curve is mirrors the unique expression of the shadow hiding effect. 
  
The goal of this paper is to understand the role played by the two known opposition effects 
(coherent backscatter and shadow hiding) on the morphology of the surge for different surface materials, 
which have different values of grain size, regolith grain filling factor, 
absorption factor (or inverse albedo), particle filling factor and vertical 
extension, by making some comparisons with the three morphological output parameters A, HWHM and $S$.

This paper describes the results of a full morphological parametrization and 
comparison of phase functions of the main satellites and rings of the Solar System 
in order to compare the influence of parameters not yet implemented in 
actual models and simulations. 
Section~2 describes the data set that we used here and the specific 
reduction we added to these previously published data in order to compare 
them more easily. We also present the morphological models that have supplied 
the parameters and discuss their link with physical properties of 
the surfaces. In Section~3, we focus on the specific behaviors of the 
morphological parameters, as a function of the single-scattering 
albedo, the distance from the Sun and the distance from the center of the
parent planet. Section~4 is dedicated to a discussion in which we physically interpret  
the general behaviors obtained with a deconvolution method. Conclusions and future work that would be of interest are drawn in Sections~5.
 
%\newpage
\section{Data set description and reduction}
\subsection{The opposition effect around a selection of rings and satellites of the giant planets} \label{oe_everywhere}
We have applied a fitting procedure to a set of phase curves of satellites 
and rings obtained by previous ground-based and \textit{in situ} optical observations 
(see Table~1 for references). The spectral resolution of the filters used for these observations are not rigorously mentionned by theirs authors, and because we mix for some objects phase curves of close wavelength, we give an approximate value of the wavelength of observation (the uncertainty of the approximated values is roughly 100 to 200~nanometers). \\
For a comprehensive study of the morphology of the opposition phase curves, 
the solar phase curves of the Galilean satellites (Io, Europa, Ganymede and 
Callisto) and the jovian main ring were chosen, as well as the phase curves of the 
Saturnian rings~(the classical A,~B, and C rings and the tenuous E~ring) and some
Saturnian satellites (Enceladus, Rhea, Iapetus and Phoebe); the rings and satellites 
of Uranus [We refer to the seven innermost satellites of Uranus -- Bianca, Cressida, Desdemona, Juliet, Portia, 
Rosalind and Belinda -- as the Portia group, to follow the designation 
of Karkoschka~(2001). The phase function of the Portia group is then the averaged phase function for
these seven satellites.], including the Portia group and three other Uranian satellites, 
Titania, Oberon and Miranda; and finally two Neptunian ring arcs (Egalit\'e and 
Fraternit\'e) and two satellites of Neptune (Nereid and Triton). For all the satellites of this study, the phase function is representative of the leading 
side because they have, in general, better coverage at small phase angles (except for Iapetus which have a trailing side brighter, we then use the Iapetus' trailing side data). References for the phase 
curves that we use in this study are given in table~\ref{tab_phase_curves_assbs}.

\textbf{Insert Table~\ref{tab_phase_curves_assbs}}

This study should give an extensive comparison between rings around the giant planets 
(Jupiter, Saturn, Uranus and Neptune), as well as a comparison between rings and 
satellites for each giant planet of our Solar System. For practical purposes, 
the well-known phase curve of the Moon is added as a reference.

\subsection{Data set reduction}
In order to  properly compare the morphological parameters of the objects whose phase curves are given as magnitudes, we have converted the magnitude~$M$ to the disk-integrated brighness~$I/F$ by using~:

\begin{equation}
I/F=10^{-0.4M}
\end{equation}

\citep{1995Icar..115..228D}. This modification allows us to directly compare the 
slope of the linear part of all the curves in the same unit. 

%\newpage
\subsection{Data set fits: the morphological models}
The purpose of the present paper is to provide an accurate description of the morphological behavior 
of the observed phase curves. This is the very first step prior to any attempt to perform
either analytical or numerical modeling. As a consequence, special care has been given here to 
parametrizing the observations efficiently and conveniently. In addition, morphological 
parametrization is necessary to efficiently compare  numerous phase curves and derive 
statistical behavior, as will be done in  Section\,\ref{results}.  \\
Several morphological models have been used in the past to  quantitatively describe the 
shape of the phase functions~: the logarithmic model of \citet{1970sips.conf..376B}, 
the linear-by-parts model of \citet{1976AJ.....81..865L} and the linear-exponential 
model of \citet{2001JQSRT..70..529K}. The specific properties of these three models 
make them adapted for different and complementary purposes. The logarithmic model 
is an appropriate and simple representation of the data, the linear-by-parts model 
is convenient to describe the shape in an intuitive way, and finally the linear-exponential 
model is commonly used for the phase curves of Solar System bodies  
\citep[see the comparative study of][]{2001JQSRT..70..529K}.

\begin{figure}[!ht]
\begin{center}
\includegraphics[width=13cm]{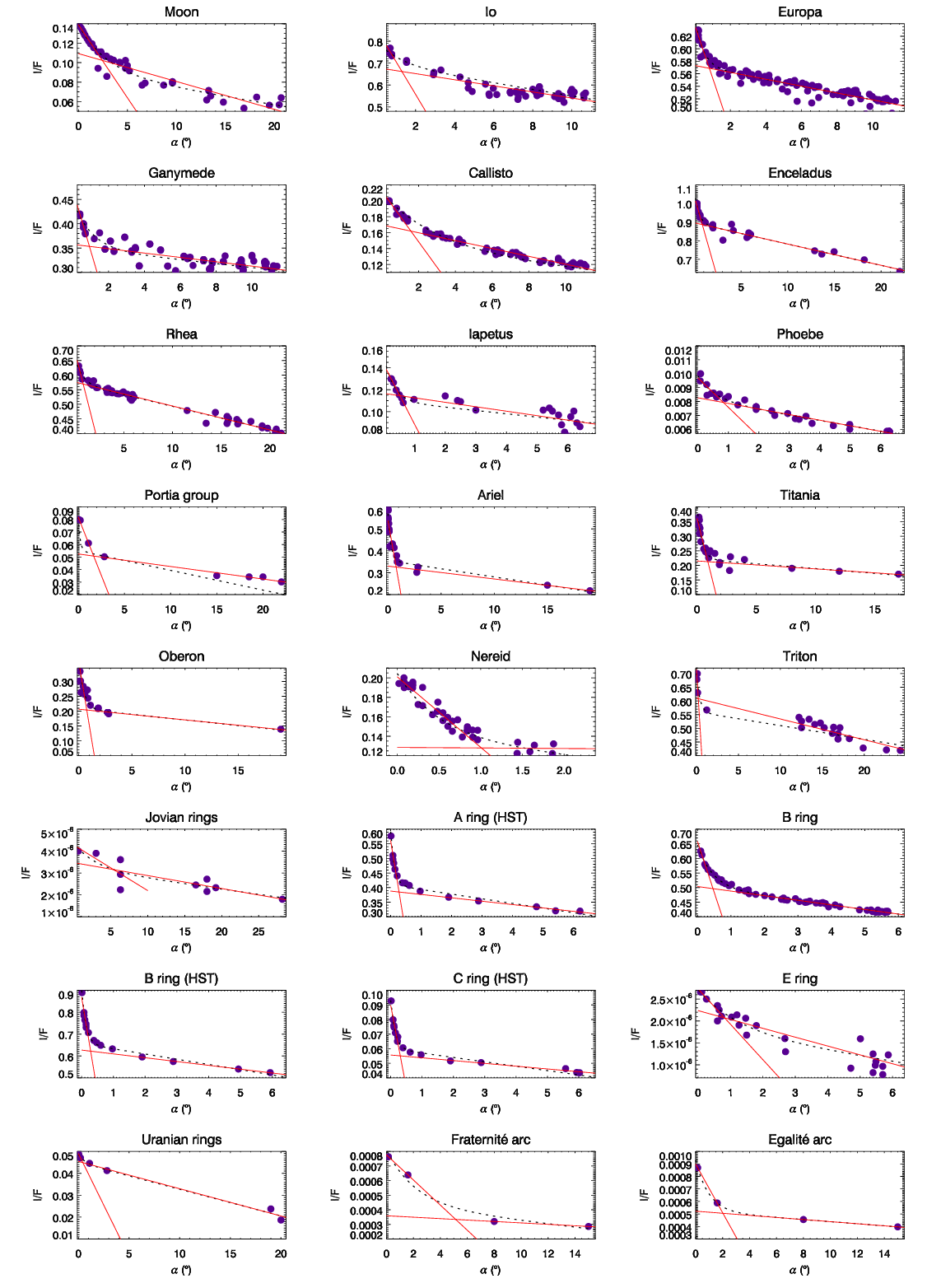}
\newline
\caption{\label{fig1} Phase curves of a selection of rings and satellites in
the Solar System (see Table~1 for
references). The solid curves correspond to the best fit obtained with
the linear-by-parts model and the dotted curves to the best
linear-exponential fit.}
\end{center}
\end{figure}

\textbf{Insert Fig.~\ref{fig1}}
 
\subsubsection{The linear-by-parts model} 
For an intuitive description of the main features of the phase curves, the linear-by-parts 
model is the most convenient one. It is constituted of two linear functions fitting both 
the surge at small phase angles ($\alpha<\alpha_{1}$) and the linear regime at larger phase angles 
($\alpha>\alpha_{2}$), where, generally, $\alpha_1 \neq \alpha_2$. Besides $\alpha_1$ and $\alpha_2$, this function depends on 4~parameters, $A_0$, $B_0$, $A_1$, and $B_1$, such that:

\begin{eqnarray}
I/F(\alpha<\alpha_{1})=-A_{0}\cdot\alpha + B_{0} \\
I/F(\alpha>\alpha_{2})=-A_{1}\cdot\alpha + B_{1}
\end{eqnarray}

\citet{1976AJ.....81..865L} and \citet{1979AJ.....84.1408E} use 
$\alpha_{1}$=0.27\textsuperscript{o} and $\alpha_{2}$=1.5\textsuperscript{o}. 
By testing several values of $\alpha_{1}$, it appears that for our data set, 
values of $\alpha_{1}=0.5$\textsuperscript{o} and $\alpha_{2}$=2\textsuperscript{o} 
provide the best results, so these values are now adopted in the rest of the paper except for the Moon and tenuous rings for which we take $\alpha_{1}=1$\textsuperscript{o}. \\ 
In terms of the four parameters $A_{0}$, $B_{0}$, $A_{1}$, and $B_{1}$, the shape of the curve 
is characterized by introducing three morphological parameters~: A, HWHM and S 
designating the amplitude of the surge, the half-width at half-maximum of the surge, and the absolute slope
at ``large'' phase angles (i.e., a few degrees up to tens of degrees), respectively. The parameters are defined by~:

\begin{equation}
\textrm{A}=\frac{B_{0}}{B_{1}}~~~~~~~~\textrm{HWHM}=\frac{(B_{0}-B_{1})}{2(A_{0}-A_{1})}
~~~~\textrm{and}~~~~ S=A_{1}
\end{equation}

Even if all the part of the opposition curve cannot be fitted by two linear functions, this model offer a convenient description of the main trends of the phase curve.

\subsubsection{The linear-exponential model} 
The linear-exponential model describes the shape of the phase function as a 
combination of an exponential peak and a linear part. Its main interest is 
that it has been used in previous work for the study of the backscattering part of the phase curves 
of the Solar System's icy satellites and rings \citep{2001JQSRT..70..529K,2002Icar..158..224P}. \\
However, as noted by \citet{2007PASP..119..623F}, we find that this model 
does not fit the phase curves well: in particular A, HWHM and S are under-
or overestimated. In addition, the converging solutions found by a downhill 
simplex technique have large error bars, which means that a large set of solutions 
is possible and thus produce some difficulties for the comparison with the other
objects. \\ 
For completeness, we give the four parameters of this model~: the intensity of the peak 
$I_{p}$, the intensity of the background $I_{b}$, the slope of the linear part $I_{s}$ and the angular  width of the peak $w$ such that the phase function is represented by~:

\begin{equation}
I/F=I_{b} + I_{s}\cdot\alpha + I_{p}\cdot
e^{-\frac{\alpha}{2w}}
\end{equation}

As $\alpha \rightarrow 0$, $\exp({-\alpha/2w}) \rightarrow 1 -\alpha/2w + \mathcal{O}$($\alpha^2$), so that the slope approaches $I_{s} + I_{p}/(2w)$. The degeneracy of these parameters may explain some of the difficulty in obtaining good fits described above. 
For consistency with previous work, we can express the amplitude and HWHM of the opposition surge in this model as: 

\begin{equation}
\textrm{A}=\frac{I_{p}+I_{b}}{I_{b}}~~~~~~~~\textrm{HWHM}=2\cdot
\ln2w~~~~\textrm{and}~~~~S=-I_{s}
\end{equation}

We report in Table\,\ref{tab_ahwhms_assbs}, the morphological parameters of the linear-by-parts model and that of the linear-exponential model.

\textbf{Insert Table~\ref{tab_ahwhms_assbs}}

\subsubsection{The logarithmic model} \label{bobrov}
As noted by \citet{1970sips.conf..376B}, \citet{1976AJ.....81..865L} and 
\citet{1979AJ.....84.1408E}, we remark that a logarithmic model 
describes the phase curves very well. It depends on two parameters ($a_{0}$ and $a_{1}$). This model has the following form~:

\begin{equation}
I/F=a_{0} + a_{1}\cdot\ln(\alpha)
\end{equation}

In general, this model is the best morphological fit to the data. However, $a_{0}$ and $a_{1}$ are not easily expressed in terms of A, HWHM and S, since the model's dependence on $\alpha$ is scale-free. Thus we report the values of
these two parameters in table~\ref{tab_all_params_assbs} to allow an easier reproduction of the observational data.

\subsubsection{A method that takes into account the angular size of the Sun}  \label{convol_method}
For all the phase curves presented here, a comparison of their surges could be compromised 
because they have different values of their observed minimum phase angle values. For example, data for the Galilean satellites never reach 0.1\textsuperscript{o}, whereas data for Saturn's rings almost reach 0.01\textsuperscript{o}.   \\
Although the behavior within the angular radius of the Sun represents a small part of the phase function, these smallest phase angles 
are crucial to constrain the fit, especially for the linear-exponential model. Indeed, when $\alpha \rightarrow 0$, the linear-exponential function tends toward $I_{b}+ I_{p}/(2w)$; as a consequence, this function flattens at very small phase angles. \\ 
However, in some cases this flattening does not correspond to the expected flattening due to the angular size of the Sun because the linear-exponential flattening fits itself arbitrarily with the phase angle coverage. The less points there are at small phase angles, the sooner will occur the flattening of the phase function. \\
\citet{2008Icar..666..666D} showed for Saturn's rings that the behavior of the surge was accurately represented by a logarithmic model between 15\textsuperscript{o} and 0.029\textsuperscript{o}, where 0.029\textsuperscript{o} corresponds to the angular size of the Sun at the time of the Cassini observations. Below 0.029\textsuperscript{o}, the resulting phase function flattens, whereas the logarithmic function continues increasing. The use of this fact observed specifically for Saturn's rings and its generalization to the Solar System objects of this study allows us to create extrapolated data points. Indeed, it is more convenient to use extrapolated data points than convolve the linear-exponential function or the logarithmic function with the solar limb darkening. First because, for incomplete phase functions, the flattening of linear-exponential function is almost uncontrollable and second because for the logarithmic model, even if a convolution is possible, linking the morphological parameters A, HWHM and S to the outputs $a_0$ and $a_1$ is not trivial. \\
The method to create extrapolated data points consists of first fitting the logarithmic model to the data and then taking the value of the logarithmic function at the phase angle which corresponds to the solar angular size ($\alpha_{\odot}$, see Appendix). We then give to six points the same $y$-value~: I/F($\alpha=\alpha_{\odot}$) and $x$-values ranging from 0.001\textsuperscript{o} to $\alpha_{\odot}$ of phase angle. These extrapolated data points are represented in Figure\,\ref{fig2} (the full method is detailed in the Appendix). The extrapolated data and the original data are then fitted by the linear-exponential model in the last step.

\textbf{Insert Fig.~\ref{fig2}}

For the Moon, Ariel and Oberon, for which the phase curve has a few points below the solar angular radius, we can see that the extrapolated points match the observational points quite well. This proves that the solar angular size effect is a flattening of the phase function below $\alpha_{\odot}$. In the case of the HST data for Saturn's rings, for which we have also a few points below the solar angular radius, the extrapolated points are a bit smaller than the observed points. This is may be due to the fact that the data of \citet{2007PASP..119..623F} are already deconvolved by another method. 

We also performed a convolution of the linear-exponential function to a limb darkening function (see Appendix), but this refinement did not significantly change the values of A, HWHM and S, because the linear-exponential model already flattens as $\alpha \rightarrow 0$. Thus, by adding extrapolated data below $\alpha_{\odot}$, we are sure that the resulting fitting function will have a constant behavior below $\alpha_{\odot}$ and that the resulting fitting function will take into account the angular size of the Sun. However, we assume that all bodies have a logarithmic increase up to the solar angular radius, which is only confirmed for Saturn's rings. Output parameters of our best fit for the ``extrapolated linear-exponential'' function are given in table\,\ref{tab_all_params_assbs}.

\textbf{Insert Table~\ref{tab_all_params_assbs}}
%\newpage

\subsubsection{A method of solar size deconvolution}  \label{deconvol_method}
Although the behavior at phase angles smaller than the solar angular radius represents a small part of the surge, a comparison of the
surge of Solar System rings and satellites could be compromised 
because they have different values for the mean solar angular radius 
($\alpha_{\textrm{min}}$=0.051, 0.028, 0.014, 0.009\textsuperscript{o} 
respectively for Jupiter, Saturn, Uranus and Neptune at their mean distances from the Sun). 
Indeed, according to the results presented here, the amplitude and HWHM seems linked to the finite angular size of the Sun. 
However, our morphological study doesn't clearly show that the Sun's angular size effect is preponderant for the amplitude~A, 
because even considering more accurately the ``solar environmental effect,'' the surges of Neptune's satellites have 
smaller amplitudes than those of Uranus (figure\,\ref{fig7}b). 
This contradicts the theoretical assumption 
that the solar angular size would give the largest amplitude to the most distant objects, for which the Sun has the smallest angular size. 
Because we previously noted that the effect of the ``solar environmental effect'' was to flatten 
the phase function when the phase angle is less than or equal to the solar angular radius 
\citep{2008Icar..666..666D}, a naive deconvolution method 
would be to allow the phase curve to rise below $\alpha_{\odot}$. This is also suggested by a previous 
deconvolution of HST data on Saturn's rings, for which the brightness still increases below $\alpha_{\odot}$ \citep{2007PASP..119..623F}. However, the 
linear-exponential function is not appropriate for this purpose because it intrinsically flattens as $\alpha \rightarrow 0$. 
Thus the only morphological function that allows an increase, even at very small phase angles, 
is the logarithmic function.  In particular, this function allows the same increase of the brightness 
above and below the solar angular radius (without break in the brightness), then using this function simulates a point source of light. However, we assume that the 
smallest phase angles are about $\alpha$=0.001\textsuperscript{o}, in this way, if a physical flattening should be performed by the coherent backscattering or the shadow hiding effect, it will be possible at these phase angles. \\
The logarithmic function fits the phase function quite well at 
small and large phase angles \citep[0.1-15\textsuperscript{o},][]{2008Icar..666..666D}; 
however, none of the morphological 
parameters A, HWHM and S are well-defined in this model. Because the logarithmic model is a good representation of the data and perform a kind of deconvolution at 
phase angles less than $\alpha_{\odot}$, we fit this function by the linear-by-parts function. 
As shown in figure\,\ref{fig9}, the fitting results of this crude deconvolution method is reasonably 
acceptable when the phase function is plotted on a linear scale of phase angle. 

\textbf{Insert Fig.~\ref{fig9}}

However, on a logarithmic scale of $\alpha$, the linear-by-parts fit is obviously 
not acceptable because intrinsically, a linear function cannot fit a logarithmic increase. As a consequence, we have slightly changed the linear-by-parts parameters in order to take into account the inappropriate flattening of this function, compared to the logarithmic function. Because the y-intercept~$B_{0}$ of the linear-by-parts model is less than the values of the logarithmic function when $\alpha$<0.01\textsuperscript{o}, we replace $B_{0}$ by the value of the logarithmic function when $\alpha$=0.001\textsuperscript{o}~:

\begin{equation}
B_{0}'=a_{0} + a_{1}\cdot\ln(10^{-3}), 
\end{equation}

where $a_1 \leq 0$, in such a way that the amplitude and the angular width are now given by~:

\begin{equation}
A=\frac{B_{0}'}{B_{1}}~~~~~~~~\textrm{and}~~~~~~~~\textrm{HWHM}=\frac{(B_{0}'-B_{1})}{2(|a_{1}|-A_{1})}
\end{equation}

where $|a_{1}|$ is the absolute slope of the logarithmic function (see section\,\ref{bobrov}). 
Replacing $|a_{1}|$ by $A_{0}$ in the formula for HWHM was motivated by the fact that the original values of the linear-by-parts parameters were systematically the same (HWHM$\thicksim$0.22\textsuperscript{o} for $\alpha_{1}$=0.3\textsuperscript{o}). This is due to the fact that the logarithmic function is a fractal function, so it is not possible to obtain a Half-Width at Half Maximum. Values of the linear-by-parts parameters A, HWHM and S are given in Table\,\ref{tab_ahwhms_ideal_assbs}. Our best result is probably that for the B~ring, for which we previously found different values of A from the \citet{1965AJ.....70..704F} data and the \citet{2007PASP..119..623F} data (A=1.30 and A=1.38 respectively, see table\,\ref{tab_ahwhms_assbs} with the convolved models). The discrepancy was still present with the extrapolated linear-exponential model (A=1.35 for \citet{1965AJ.....70..704F} and A=1.32 for \citet{2007PASP..119..623F}), due to the fact that the solar angular size was different at the two observation times (see table~Appendix\ref{tab_sunsize_assbs}). Now, with the unconvolved model, the discrepancy of the two values is somewhat reduced compared to values from the fit to the original data~: A=1.82 for \citet{1965AJ.....70..704F} and A=1.77 for \citet{2007PASP..119..623F}, table\,\ref{tab_ahwhms_ideal_assbs}, which implies that the angular size effect is now absent. 

\textbf{Insert Table~\ref{tab_ahwhms_ideal_assbs}}

%\newpage
\section{Results} \label{results}

Our procedure is to interpret more carefully the morphological results of a large dataset. We start first by studying the behaviors of the morphological parameters with the single scattering albedo with the raw data (Section\,\ref{results_1} and \ref{results_2}) and the improved data that take into account of the solar size of the Sun (Section\,\ref{results_3}). In a last step, we freed from the solar size biais by trying to look the opposition effect in the outer Solar System with the same solar size, assumed to be a point (Section\,\ref{results_4}).

\subsection{Behaviors of the morphological parameters} \label{results_1}
In this section we compare the morphological parameters as a function of the 
single scattering albedo~$\varpi_{0}$. \\
Since the single scattering albedo~$\varpi_{0}$ represents the ratio of scattering efficiency to total light extinction over the phase angle range \citep{1960ratr.book.....C}, its value must be computed with the largest coverage of phase angle as possible (0 to 180 degrees). This is why we did not compute the single scattering albedo with the phase curves 
presented in this paper but we use previously published values of single scattering albedo computed from phase curves with a larger phase angle coverage than ours and a wavelength close to ours. Thus, references for phase curves (table~\ref{tab_phase_curves_assbs}) and 
references for $\varpi_{0}$ (table~\ref{tab_albedo_assbs}) are not always the same.   

\textbf{Insert Table~\ref{tab_albedo_assbs}}

We did not found single scattering albedo values for the jovian main ring and the Saturn's E ring, so these two objects will be excluded of the study of the mophological parameters with the single scatterig albedo.

\subsubsection{Variation of the angular width of the surge with albedo} 
First, we discuss the variation of HWHM=$f(\varpi_{0})$ derived from the linear-by-parts model (Figure~\ref{fig3}a) and HWHM=$f(\varpi_{0})$ derived from the extrapolated linear-exponential model (Figure~\ref{fig3}b). Interestingly, the variation differs according to the morphological model~: the first case leads to a decrease of HWHM when $\varpi_{0}$ increases, while the latter case leads to an increase of HWHM when $\varpi_{0}$ increases. 

\textbf{Insert Fig.~\ref{fig3}}

The different results for HWHM=$f(\varpi_{0})$ between the linear-by-parts results (well fitted by HWHM$\thicksim0.9\times0.3^{\varpi_{0}}$, Figure\,\ref{fig3}a) and the extrapolated linear-exponential results (represented by HWHM$\thicksim0.15+0.19\varpi_{0}$, Figure\,\ref{fig3}b) is mainly due to the points which correspond to the Moon, Callisto and Nereid. Indeed, in general, values from the extrapolated linear-exponential model significantly decrease for the outer Solar System objects, whereas the value for the Moon increases by almost 1\textsuperscript{o}. This is due to the fact that when we take into account the Sun's angular size, this effect lower the values of HWHM for the incomplete phase functions.  \\
However, Figure~\ref{fig3} shows a large dispersion of HWHM with albedo. 
 
%\newpage 
\subsubsection{Variation of the amplitude of the surge with albedo} 
Figure~\ref{fig4} shows a weak dependence of the amplitude of the surge on the 
albedo for the satellites, already noted by \citet{1997Icar..128....2H,2002ocd.book..224}. 

\textbf{Insert Fig.~\ref{fig4}}

In both cases (linear-by-parts model, Figure~\ref{fig4}a and extrapolated linear-exponential model, 
Figure~\ref{fig4}b) we note a dependence of $A$ with $\varpi_{0}$, 
which follow a function leading 
to a decrease of $A$ when $\varpi_{0}$ increases (A$\thicksim1.65\times0.72^{\varpi_{0}}$ in Figure\,\ref{fig4}a and A$\thicksim1.75\times0.72^{\varpi_{0}}$ in Figure\,\ref{fig4}b). The consistent trends in both cases imply that the finite size of the Sun was correctly derived by the linear-by-parts model. \\
The decrease of $A$ with increasing $\varpi_{0}$ could be understood 
by a relation between the amplitude and the single scattering albedo via the 
intensity of the background phase function~$I_{b}$ (with the linear-exponential model), 
which is inversely proportional to the albedo. Thus, the predicted trend of \citet{1990AdSpR..10..187L} 
is confirmed by our present results. \\
In addition, Figure\,\ref{fig4}b labels satellites by color to indicate their parent planet. This figure indicates that distant objects (such as the Uranian satellites) have a significantly larger amplitude than less distant objects (such as the Galilean or Saturnian satellites) while the Neptunian satellites have values in the average. Then it must be considered 
that the finite size of the Sun has a role in the amplitude's value \citep{1991AVest..25...71S}.
As a consequence, even if the trend of A=$f(\varpi_{0})$ is well explained by theoretical considerations, one can 
remark that the large dispersion in this correlation could be due to other effects (such as the finite size of the Sun) that weakens the albedo dependence of A. Thus, as for HWHM, we cannot physically interpret
the variation of HWHM and A as long as they are convolved with the effect of the solar angular size. 
 
%\newpage
\subsubsection{Variation of the slope of the linear part with albedo}
The last morphological parameter is the slope S, which we represent as a function of the single
scattering albedo $\varpi_{0}$ for the rings (Figure~\ref{fig5}a) and satellites (Figure~\ref{fig5}b) of the Solar System. 

\textbf{Insert Fig.~\ref{fig5}} 

In this figure, rings and satellites have different values of 
slope as function of their albedo, and a slight increase for S with increasing 
$\varpi_{0}$ is noticed. For the rings (Figure~\ref{fig5}a), it seems that a good
correlation appears between S and the albedo, which may be roughly fitted by a function like 
S$\thicksim$0.001+0.02$\cdot\varpi_{0}^{2}$. A similar fit works well for 
the satellites (the Moon, Saturnian and Uranian satellites are not far from the dashed line in
figure~\ref{fig5}b). This fit to the points could be S$\thicksim$0.001+0.01$\cdot\varpi_{0}^{2}$ (Figure~\ref{fig5}b). 
However, three objects fall far from this curve~: Europa, Ganymede and Io. This correlation 
suggests that multiple scattering may be a strong element at play in the regime of 
self-shadowing (beyond $\thicksim$1\textsuperscript{o} of phase angle), in qualitative 
agreement with \citet{1974IAUS...65..441K}. 

%\newpage
\subsection{Cross comparisons between the morphological parameters of the surge}  \label{results_2}
We see in Figure\,\ref{fig3} that the angular width of the surge can change significantly by taking 
into account the solar angular radius. However, it is not the case for the amplitude of the surge (Figure\,\ref{fig4}).   
Figure~\ref{fig6} shows the behavior of a cross comparison between the morphological 
parameters A and HWHM obtained with the linear-exponential model convolved  (Figure~\ref{fig6}b) or not (Figure~\ref{fig6}a) with the limb darkening function. 

\textbf{Insert Fig.~\ref{fig6}}

In the first graph (Figure~\ref{fig6}a), it first seems that two different groups may be qualitatively distinguished. \\
On the one hand, there is a group of objects with similar values of the HWHM, in the 
range 0.1\textsuperscript{o} to 0.4\textsuperscript{o}, but with significantly 
different values of the amplitude, from 1.4 to 1.8. It is interesting to note that 
these bodies, which include the Saturnian rings and Uranian satellites, are not bodies in 
the outermost part of the Solar System (such as the Neptunian satellites). 
Within this group, we also note that similar objects are gathered in the (A,HWHM) space: the 
Uranian satellites have, on average, the largest values of the amplitude, $\thicksim$1.7.  
Saturn's rings have an amplitude between 1.3 and 1.6, closer to the Uranian 
satellites. We also note that whereas all satellites have quite a
constant HWHM (between 0.2\textsuperscript{o} and 0.4\textsuperscript{o}), Saturn's rings 
have systematically lower values, between 0.08\textsuperscript{o} and 0.09\textsuperscript{o},
which may be suggestive of a different state of their surface. \\
The second group includes Saturn's satellites, along with Io, Europa and Triton, which 
have the lowest values of amplitude. A very striking feature is the peculiar behavior 
of bodies such as Callisto and the Moon~: they have similar amplitudes 
(about 1.5) and also similar HWHMs (about 2\textsuperscript{o}). 
In the second graph (Figure~\ref{fig6}b), it first seems that the bodies belonging to the same primary planet have similar values of A and HWHM. For the Galilean satellites, we found the largest HWHM for the outer Solar System satellites (between 0.2\textsuperscript{o} and 0.5\textsuperscript{o}) 
and amplitude between 1.1 and 1.5. The Uranian satellites still have the largest amplitudes (A ranges between 1.6 and 1.9), but the values of HWHM are similar to that of those of Saturn's rings and satellites. The Neptunian satellites have the sharpest opposition peaks (HWHM$\lesssim$0.1\textsuperscript{o}) but moderate amplitudes (between 1.2 and 1.4), similar to the range of the Saturnian satellites. \\
Does this imply some deep structural difference of the surface 
regolith of bodies, or is it due to the Sun's angular size effect? For
the moment we note that the opposition effect is poorly understood,
especially at phase angles smaller than 1\textsuperscript{o} in the
coherent backscattering regime. \\
Whereas physical implications are still hard to draw from these graphs, 
it is interesting to note that the solar angular size refinement that we use naturally 
clusters different kind of surfaces in different locations of the (A, HWHM) space, and that
``endogenically linked objects" are quite well gathered in small portions of this space. 
This could suggest that common environmental processes 
(meteoroid bombardment, surface collisions, space weathering, etc.) 
may homogenize different surface states by processing mechanisms that may 
determine the microstructure of the surface, and then, 
in turn, the behavior of the opposition surge at very low phase angles, 
as it may be linked with the spatial organization of micrometer-scale
surface regolith \citep{1992Ap&SS.194..327M,1992MNRAS.254P..15M,1999Icar..141..132S}. 

\subsection{Additionnal effect} \label{results_3}
With the cross comparison of the morphological parameters of the surge, the angular size of the Sun and the fact that objects seem ``endogenically linked", two supplementary effects (observationnal and physical) can significantly modify the values of A and HWHM~: the ``solar size bias'' and the ``planetary environmental effect''. 
\subsubsection{The ``solar size bias''}
 
We tested the influence of the solar angular size by representing in Figure\,\ref{fig7} the morphological parameters of the surge A and HWHM as a function of the distance from the Sun~$d$ (in Astronomical Units).

\textbf{Insert Fig.~\ref{fig7}}

We represent HWHM (Figure\,\ref{fig7}a) and A (Figure\,\ref{fig7}b) from the linear-by-parts model with filled symbols and that of the extrapolated linear-exponential model with empty symbols. \\
We remark that the linear-by-parts angular width follows the power-law function~HWHM$\thicksim$0.33+1.1$d^{-1.5}$ (the solid line in Figure\,\ref{fig7}a). The fit is quite good from the Moon to Uranus, but is far from the values of Neptune's satellites (especially that of Nereid). It is easier to see with this representation that the HWHM of Nereid is larger than that expected by the power-law function. The extrapolated linear-exponential HWHM corrects this because the Neptunian satellites now have smaller values of HWHM that are better fitted by the power-law function. Indeed, the extrapolated linear-exponential HWHM follows a similar function (HWHM$\thicksim$0.12+2.3$d^{-1.4}$), but values at the extreme parts of the Solar System (innermost with the Earth's satellite and outermost with Neptune's satellites) are significantly different~: for the Moon, the extrapolated linear-exponential HWHM is larger than its linear-by-parts counterpart and for the Nereid, the extrapolated linear-exponential HWHM is smaller than the linear-by-parts HWHM. \\
However, such a strong trend is not observed in the case of the amplitude of the surge. As shown in Figure\,\ref{fig7}b, a fit to the linear-by-parts amplitudes is good for the Galilean, Saturnian and Uranian satellites (which we fit by a linear function~A$\thicksim$1.1+0.08$d$) but not at all for the Moon and the Neptunian satellites. The predicted behavior (dashed line in Figure\,\ref{fig7}b) shows that the value for the Moon is overestimated and that the values of the Neptunian satellites are strongly underestimated. The use of values of the extrapolated linear-exponential A did not improve the fit. Indeed, the extrapolated linear-exponential amplitude is larger than the linear-by-parts amplitude for the Moon, whereas the extrapolated linear-exponential values of the Neptunian satellites are smaller than their linear-by-parts counterparts. The exact opposition trends were expected to obtain a good linear fit from the Moon to Neptune. Perhaps the solar size effect is not important at Neptune's distance and the values of A and HWHM are physical, in the sense that they depend only on the opposition effect mechanisms (coherent backscattering and shadow hiding).  \\ 
These results seems to suggest that A is less affected by the ``solar size bias'' than HWHM, which is entirely controlled by this effect (which seems trivial because this bias is an angular effect). It is possible that the values of A result from a coupling of the physical opposition effects (coherent backscatter and shadow hiding) with the environmental opposition effects (solar and planetary). As a consequence, the deconvolution of the phase function (at least for the ``solar size bias'') should allow the physical opposition effects to express fully themselves in the values of A and HWHM.  

%\newpage
\subsubsection{The ``planetary environmental effect''}
Previous studies by \citet{2006Icar..184..181B} and \citet{2007Sci...315..815V} have confirmed, 
at the scale of the Saturnian system, a kind of ``endogenic" or ``ecosystemic" classification of 
the opposition surge. Indeed, these works demonstrated that the opposition surge paramaters of 
the outermost and innermost Saturnian satellites, respectively, can be 
a function of the distance from Saturn. \\
For the planetary environments of Jupiter, Uranus and Neptune, there is no significant variations with distance from the parent planet. The first reason is maybe statistical because there are not enough data to make  (for these systems, we have less than four objects). The second is that the dust environment can be influenced by other effects (the magnetospheric activity, the satellite's activity, the proximity to the Kuiper Belt and transneptunian objects), in such way that the distance from the planet host could be irrelevant for some of the planetary environments. \\
For the Saturnian system, for which local interactions between satellites and rings exists, we observe similar trends with our results than the trends of \citet{2006Icar..184..181B} and \citet{2007Sci...315..815V} (see Figure~\ref{fig8}). 

\textbf{Insert Fig.~\ref{fig8}}

Variations of the morphological parameters on large scales of distance (for the Saturn system) 
show trends that suggest a common ground for environmental processes. 
These processes may imply different surfaces, but will be handled by the opposition 
effect in the same way by the mechanisms that determine the microstructure of the surface. 
According to theoretical models of coherent backscattering, the amplitude is related 
to the grain size and HWHM depends on the composition, distribution of grain size and the 
regolith filling factor. Thus the behavior of the opposition surge is connected to 
the spatial organization of the regolith  \citep{1992MNRAS.254P..15M,1999Icar..141..132S}. \\
Therefore, the study of the morphology of the opposition peak can highlight dynamical interactions 
between the rings, satellites and the surrounding environment through the photometry. 
These ring/satellite interactions noticed here go beyond the general dynamical interactions 
between rings and satellites (such as resonances, for example). Here these interactions involve common erosion histories on the surfaces of the rings and satellites. 
Similar values of HWHM according to the theory of \citet{1992MNRAS.254P..15M} 
can be explained by similar values of refractive index (with various values of 
grain size and filling factor), or by different values of refractive indices, 
but similar values of grain sizes and filling factor of the regolith. \\
There are two known mechanisms that can act together in order to explain the 
similarities in the values of HWHM for objects that are \textit{endogenically linked}. \\ 
\textcircled{1}~ The impacts of debris in planetary environments can change the 
chemical composition of the rings and satellites~: new elements can be directly added to the system~;
the more volatile elements can be preferentially removed and the more fragile compounds can be preferentially processed. The work of \citet{1998Icar..132....1C}, in particular, details changes in the chemical composition of Saturn's rings by meteoroid bombardment and ballistic transport. \\ 
\textcircled{2}~The second mechanism that is likely to act concerns every kind 
of collisional mechanism capable of modifying, at microscopic scales, the surface of the 
satellite's regolith (meteoroid bombardment, external collisions, disintegration in space, etc.); see \citep{1988JGR....9313776L,1992JGR....9710227C,1993JGR....98.7387C}. 
In the case of ring particles, models of erosion by ballistic transport 
have been developed and predict the destruction of micrometer-sized grains in 
dense rings ($\tau$> 1), see \citep{1983Icar...54..253I}.

%Concerning HWHM, current models link differently HWHM with medium physical properties 
%\citep{1992Ap&SS.194..327M,1992MNRAS.254P..15M,2002Icar..157..523H}). 
%But according to \citet{2007PASP..119..623F}, the morphological HWHM is 
%less than the HWHM of the shadow hiding (SH) and greater than HWHM of the 
%coherent backscatter (CB). Consequently, the morphological width have also 
%a complex relation with SH and CB angular widths. Current models can thus 
%difficultly explain the dependence of the morphological HWHM on the albedo.
%\newpage
\subsection{Study of the unconvolved opposition parameters} \label{results_4}
With the unconvolved morphological parameters obtained with the method in Section\,\ref{deconvol_method}, we are now sure that the morphological surge parameters are independent of the distance from the Sun, and thus independent of the ``solar size bias''. Indeed, in Figure\,\ref{fig10}, we represent A and HWHM derived from the linear-by-parts model which fits the logarithmic model as a function of the distance from the Sun and there is no relation with the distance, unlike in Figure\,\ref{fig7}. 

\textbf{Insert Fig.~\ref{fig10}}

Moreover, we can see that three groups can be distinguished~:
\begin{itemize} 
\item[(1)] A group with the Moon, Callisto, the C~ring, Phoebe, the classical Uranian satellites (Ariel, Titania and Oberon) and Nereid. They have large  angular widths and amplitudes (HWHM$\gtrsim$0.5\textsuperscript{o}, Figure\,\ref{fig10}a and A$\gtrsim$2.0, figure\,\ref{fig10}b). We can see that these objects are dark, with low and moderate albedos (table\,\ref{tab_albedo_assbs}). These objects are also known to be heavily cratered \citep{2001SSRv...96...55N,2003Icar..163..263Z}; thus, they don't have intrinsic resurfacing mechanisms. For the specific case of the resurfacing of the Saturn's C ring, it is known that the collisional activity of a ring is controlled by the optical depth $\tau$ \citep{1998Icar..132....1C}. The number of collisions per orbit per particle is proportional to $\tau$ \citep[in the regime of low optical depth, see][]{1988AJ.....95..925W}, and the random velocity in a ring of thickness $H$ is 
about $H\times\Omega$ (with $\Omega$ standing for the local orbital frequency). 
Since $H$ is a decreasing function of $\tau$, impact velocities are high in regions of low optical depth. As a result, particles in low optical depth regions (such as the C~ring) may suffer of resurfacing characterized by rare, but somewhat higher-speed, collisions.  
\item[(2)] A group with Io, Iapetus, Rhea and the bright Saturnian rings characterized by smaller amplitude and angular width~: A$\lesssim$1.7 and HWHM$\thicksim$0.4\textsuperscript{o} (figures\,\ref{fig10}a,b).  
\item[(3)] A group with Ganymede, Europa, Enceladus and Triton with the smallest amplitude and angular width (1.3<A<1.6, Figure\,\ref{fig10}a and 0.1<HWHM<0.3\textsuperscript{o}, Figure\,\ref{fig10}b). Interestingly for the amplitude, we can see that this group contains only the brightest surfaces of the Solar System, with a single scattering albedo close to $\varpi_{0}\thicksim$0.9  \citep[however, Bond albedo of Ganymede is quite lower, see][]{1981Icar...46..137S}. These objects are also known to have active resurfacing. Indeed, this was confirmed for Europa, which has a very young surface and perhaps recent geyser-like or volcanic activity \citep{1998Natur.391..371S,1998Natur.391..365P}, and Ganymede, on which the grooved terrains could have formed through tectonism, probably combined with icy volcanism \citep{1998Icar..135..276P,2001Sci...292.1523M}. Present-day resurfacing is also taking place on Enceladus, whose geysers produce the E~ring \citep{2006Sci...311.1393P}, and for Triton, which also has geysers \citep{1995netr.conf..879C}. 
\end{itemize} 
This classification seems to suggest that the darkest and oldest surfaces have the largest amplitudes for the surge and that the brightest and youngest surfaces have the smallest amplitudes. However, one might be surprised that the third group does not include Io, which has an intense resurfacing via tidally induced volcanism. Also, the Portia group does not belong to only one group in figure~\ref{fig10}~: it belongs to the group~3 for HWHM and to the group~2 for the amplitude~A. For these two isolated cases, it is possible that the average of photometric phase curves from different satellites is responsible of the fact that Io and the Portia group are difficult to classify.

%\newpage
\section{Discussion}  \label{discussion} 

\subsection{Implications of the surge parameters of the unconvolved data}
By removing the ``solar size bias'', we can try to physically interpret the amplitude variations with the single scattering alebdo with the mechanisms proposed to explain the opposition effect. Our study shows a link between the single scattering albedo and the unconvolved morphological parameters. A linear fit to the unconvolved amplitude is A=$2.2 - 0.5\varpi_{0}$ (with a correlation coefficient of -61\%) and the linear fit to the unconvolved angular width is HWHM=$0.52 - 0.19\varpi_{0}$ (with a correlation coefficient of -38\%). By excluding the Portia group, we find a better correlation coefficient for HWHM~: -66\%. These correlations are stronger than that previously found with the convolved data. This shows that the ``solar size bias'' acts to scatter the morphological parameters. As a consequence, the fact that old and dark surfaces with a low resurfacing activity have high unconvolved amplitude whereas the bright and young surfaces with an intense resurfacing activity have low unconvolved amplitude is linked to the single scattering albedo variations of A. Indeed the single scattering albedo is a measure of the brightness of a surface. But not only: according to \citet{1999Icar..141..132S}, the amplitude of the coherent backscattering opposition surge is a decreasing function of increasing regolith grain size. 
If the morphological amplitude is due to the coherent backscattering effect 
\citep{1992Ap&SS.189..151M,M2006}, the dependence of $A=f(\varpi_{0})$ could be understood as 
a positive correlation between the grain size and the single scattering albedo. However, it is possible that 
the morphological amplitude is not only that of the coherent backscattering effect but is dominated by both effects~: coherent backscatter and shadow hiding, as underlined by \citet{2002Icar..157..523H}. 
However, here it is not possible to separate the two effects and say which effect is dominant because to separate the coherent backscatter and shadow hiding mechanisms, the polarization is required \citep{2007JQSRT.106..360M}.

%\newpage
\subsection{Implications of the slope of linear part}

The strong correlation of the slope S (in I/F.deg$^{-1}$ units) with single scattering albedo (Figure~\ref{fig5}) implies that shadow hiding is more efficient in high albedo surfaces. This trend 
was previously remarked in the opposition slope of asteroids by \citep{2000Icar..147...94B}. 
It was first interpreted by these authors as a decrease of the absolute slope with albedo, 
consistent with the analytical model of \citet{1997Icar..128....2H} which predicted that 
the amplitude of the shadow hiding must decrease with albedo. However, because the slope 
unit in \citep{2000Icar..147...94B} is magnitude (remark that the scale of the mangitude is not reversed for their graph, figure 4, as for the the other graphes that show the phase curves, figures 1 and 2), a decreasing slope in mag.deg$^{-1}$ 
corresponds to an increasing slope in I/F.deg$^{-1}$ units. As a consequence, the behavior 
of the slope as a function of the albedo for the rings, satellites and asteroids 
of the Solar System is consistent and all lead to the same idea that shadow hiding is 
reinforced at high albedo.
The correlation between slope and albedo seems to be the strongest 
trend of the opposition effect in satellites and rings of the Solar System and the 
use of more sophisticated models is needed to understand them.

The results from Figure~\ref{fig5} are in agreement with the simulations of ray-tracing, \citep{S1999}, 
which models shadow hiding in a layer of particles. These simulations show that shadow 
hiding creates a linear part in the phase function from 10 to 40 degrees  and that the absolute slope of 
the linear part becomes steeper when optical depth increases and the filling factor of the 
layer of particles increases. \\
How does albedo relate to the optical depth and the filling factor? Previous studies have 
shown that the albedo and optical depth are highly positively correlated for the rings,  \citep[see][]{1989Icar...80..104D,1991PhDT.........8C,dones1993}. For satellites, optical depth is effectively infinite; since this removes one variable, relating the slope parameter to the nature of the surface is easier for satellites than for rings. 
Thus we must consider two kinds of objects:
\begin{itemize}
\item For rings, where the optical depth is finite, variation of the slope~S will be a subtle effect involving 
both optical depth and filling factor~;
\item For satellites, which have ``solid'' surfaces, variations in slope are linked to the filling factor of each surface. If the optical depth is 
invariant for satellites, according to the model of \citet{S1999}, only variations of the 
filling factor can explain differences in the slope~S. However, the notion of filling factor is not 
well suited for satellites; indeed, a description involving a topographical roughness is more appropriate. 
\end{itemize}
We noticed that when the optical depth is finite, as for the rings, the effects of slope 
are stronger with a high albedo than for high albedo satellites. Consequently the shadow hiding effect 
for the ring is more efficient than for satellites and reflects a difference between the 
three-dimensional aspect of a layer of particles in the rings and a planetary regolith.

%\newpage
\section{Conclusion} \label{conclusion} 
The goal of this paper was to understand the role of the viewing conditions on the morphological parameters of the opposition surge and the role played by the single scattering albedo on the morphological parameters. We have use three methods to fit the data: the first one with a simple morphological model, secondly by taking more accurately into account the role of the size of the Sun in the morphological model and thirdly eliminating the role of the size of the Sin in the morphological model. \\
The results of this study allow us to highlight several facts related to the observation and the mechanisms of the opposition effect in the Solar System.
\begin{itemize} 
\item[(1)] The slope of the linear part is an increasing function of albedo. Our results are consistent 
with those of \citet{2000Icar..147...94B}, for which the slope of the phase 
function of asteroids increases when albedo increases. These results confirm the predictions from simulations of shadow hiding
for the first time.  
\item[(2)] We note that the morphological parameters of the surge (A and HWHM) are sensitive to the 
phase angle coverage, specifically to the smallest phase angles. We have extrapolated observational 
data points in order to correct the lack of data near the solar angular radius. However, this method 
needs to be improved, for example by taking directly into account of the solar angular radius 
in the linear-exponential function. We hope that future data at the smallest phase angles, will confirm the extrapolated data that we use to perform the ``extrapolated linear-exponential'' model.
\item[(3)] The amplitude and the angular width of the opposition surge are linked to the single scattering 
albedo of the surfaces, as already noted in laboratory measurements \citep{2003AA...409..765K}. Like \citet{2000Icar..147...94B}, 
we believe that the single scattering albedo is one of the key elements constraining morphological 
parameters. However before physically interpreting these results, A and HWHM need to be 
deconvolved to the ``solar size bias'' since we have a large dispersion in the relations of 
$A=f(\varpi_{0})$ and HWHM=$f(\varpi_{0})$. 
\item[(4)] By deconvolving the phase functions to the Sun's angular size effect, we showed that A and HWHM are still correlated with the albedo (with better correlation coefficients). The dependence of A and HWHM are now independent of the distance from the Sun, unlike their convolved counterparts. Indeed, values of A and HWHM from deconvolved phase functions can be classified into three groups that include a mix of bodies from the inner and the outer Solar System. This shows that 
icy and young surfaces (such as Europa, Io, Enceladus and Triton) 
have the smallest amplitudes, whereas dark and older surfaces (such as the Moon, Phoebe and the C~ring) have the largest amplitudes. 
\item[(5)] It seems that two effects (the ``solar size bias'' and the ``planetary 
environmental effect''), act together to disperse data taken from different places in the Solar System. 
Moreover, with our technique of deconvolution of phase curves, we see that the ``solar size bias'' 
can be removed from A and HWHM, because unconvolved data have A and HWHM that don't show any trend with distance from the Sun. These arguments strengthen the conclusions 
that the notion of ``ecosystem'' for a planetary environment 
can be the key element determining the opposition effect surge morphology.
\end{itemize} 

Our method cannot directly derive the physical properties obtained from 
the models. Firstly, because there is a large set of models and it seemed more convenient 
to separate the morphological models from the more physical and sophisticated ones. Secondly, 
because the various spectral resolution of our data set is not appropriate for a majority of 
physical models which need a fine spectral resolution (for example, in the coherent 
backscatter theory, HWHM is linked to the ratio of the wavelength over the free mean path of photons). In addition, the coherent backscatter can singificantly polarized the brightness of a surface \citep{1993tres.book.....H,002sael.book.....M}, so the polarized phase curves can bring crucial and complementary informations to that of the unpolarized phase curves. Consequently, more investigations need to be provided for this purpose by using color and polarized phase curves. 

For a future work, which will critically depend on the quality of the observations, first it would be interesting 
to study the phase functions of the leading and trailing faces of synchronously rotating satellite in order to test the role of the environmental effect more preciselys. Indeed, satellites are 
subject to energy fluxes from electrons, photons and magnetospheric plasma, and ion bombardment, 
which are not the same on the leading and trailing sides \citep{1988Natur.333..148B}. 
Consequently, morphological parameters might vary significantly from the 
leading side to the trailing side for the same satellites. However, the important dispersion in trailing side data \citep[see for example][]{2001JQSRT..70..529K} did not allow us to pursue this comparison. We then hope to have in the future that more accurate data of all satellites of the satellites of the Solar System. \\
Secondly, to better understand the role of the ``planetary environmental effect,'' a more relevant study would be the comparison of rings with ``ringmoons'' or small satellites 
which are in the vicinity of the rings. Several examples of such a ring/ringmoon system are 
present in the environment of each giant planet ~:
\begin{itemize} 
\item for Jupiter~: Metis and Adrastea with the main ring \citep{1987Icar...69..458S}; 
Amalthea and Thebe with the Gossamer ring \citep{1999Sci...284.1146B}~;
\item for Saturn~: Pan, Dapnhis and Atlas with the outer A~ring \citep{1981Sci...212..163S,2006AJ....132..692S}, 
Prometheus and Pandora with the F~ring \citep{1981Sci...212..163S}, and Enceladus with the E~ring (although Enceladus is the primary source of the E~ring, it is usually not called a ``ringmoon''
 \citep{2007Sci...315..815V}~;
\item for Uranus~: Cordelia and Ophelia with the $\epsilon$ ring \citep{1995AAS...186.3302F}~;
\item for Neptune~: Galatea with the Adams ring \citep{1991Sci...253..995P}.
\end{itemize} 
Unfortunately, opposition phase curves of the small satellites are not actually 
available because they require a fine spatial resolution.

\end{linenumbers}
\label{lastpage}
%\newpage
%\section*{ACKNOWLEDGEMENTS}
\ack
The authors would like to thank S. Kaasalainen and D. French for kindly providing us with some of the solar phase curves 
used here, as well as J. Burns, K. Muinonen and D. Stankevich for useful comments that improved the quality of the paper. This work 
was supported by the French Centre National de la Recherche Scientifique and the Cassini project.

%%%%%%%%%%%%%%%%%%%%%%%%%%%%%%%%%%%%%%%%%%%%%%%%%%%%%%%%%%%%%%%%%
% Bibliography                          %
%%%%%%%%%%%%%%%%%%%%%%%%%%%%%%%%%%%%%%%%%%%%%%%%%%%%%%%%%%%%%%%%%
\newpage
\bibliographystyle{elsart-harv}
\bibliography{OE1BIS}
%\newpage
%\listoffigures

%%%%%%%%%%%%%%%%%%%%%%%%%%%%%%%%%%%%%%%%%
%   Figures
%%%%%%%%%%%%%%%%%%%%%%%%%%%%%%%%%%%%%%
\newpage

\newpage
\begin{figure}[!ht]
\begin{center}
\includegraphics[width=13cm]{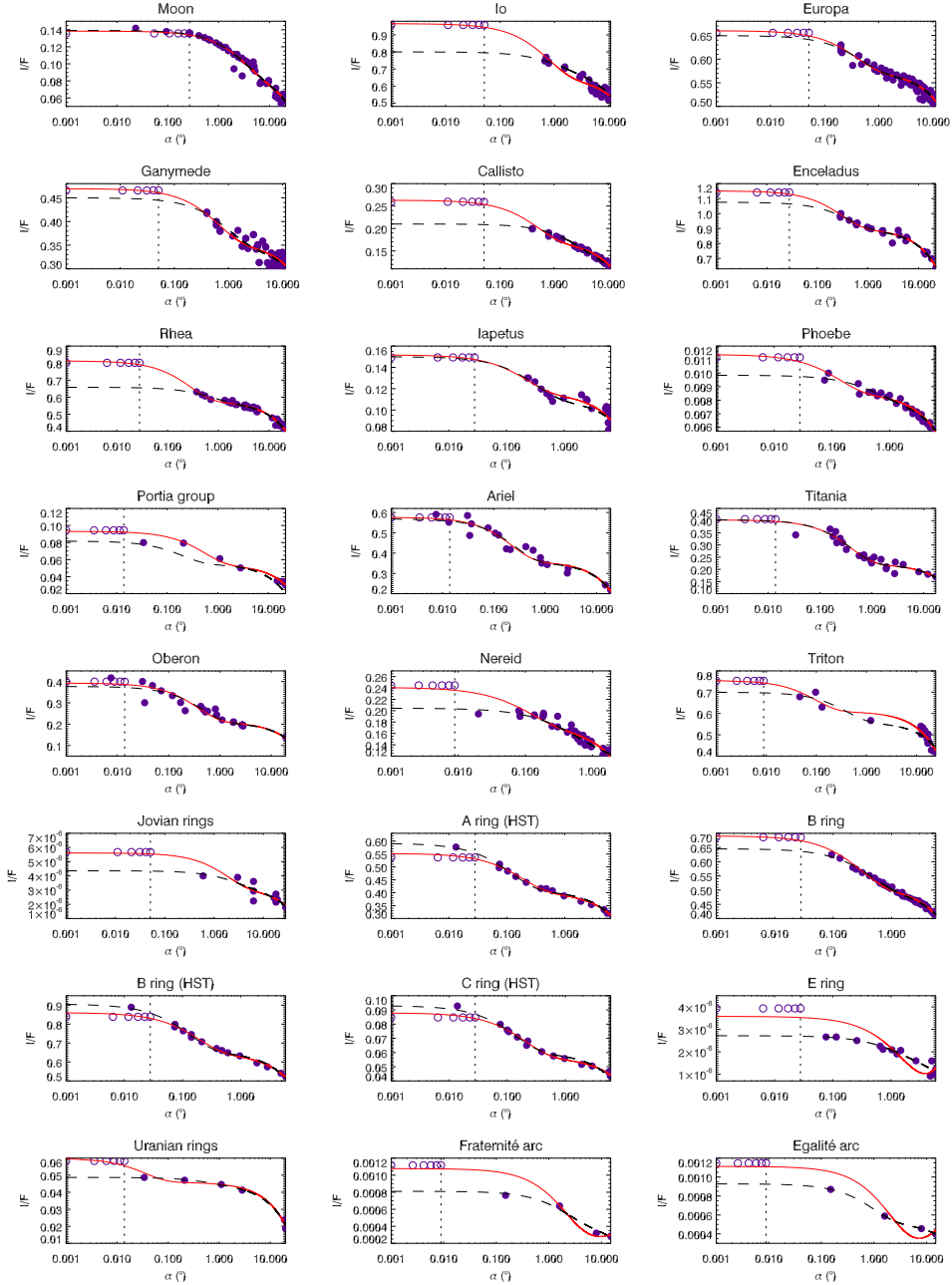}
\newline
\caption{\label{fig2} Phase curves of a selection of rings and satellites in
the Solar System. The solid curves correspond to the best fit obtained with
the linear-exponential model convolved with the size of the Sun (using extrapolated data below the angular size of the Sun, empty symbols) and the dashed curves correspond to the best linear-exponential fit (using only original data, filled symbols). 
The vertical dotted lines represent the angular size of the Sun at the observation time (see Table\,\ref{tab_sunsize_assbs} of Appendix).}
\end{center}
\end{figure}

\newpage
\begin{figure}[!ht]
\begin{center}
\includegraphics[width=13cm]{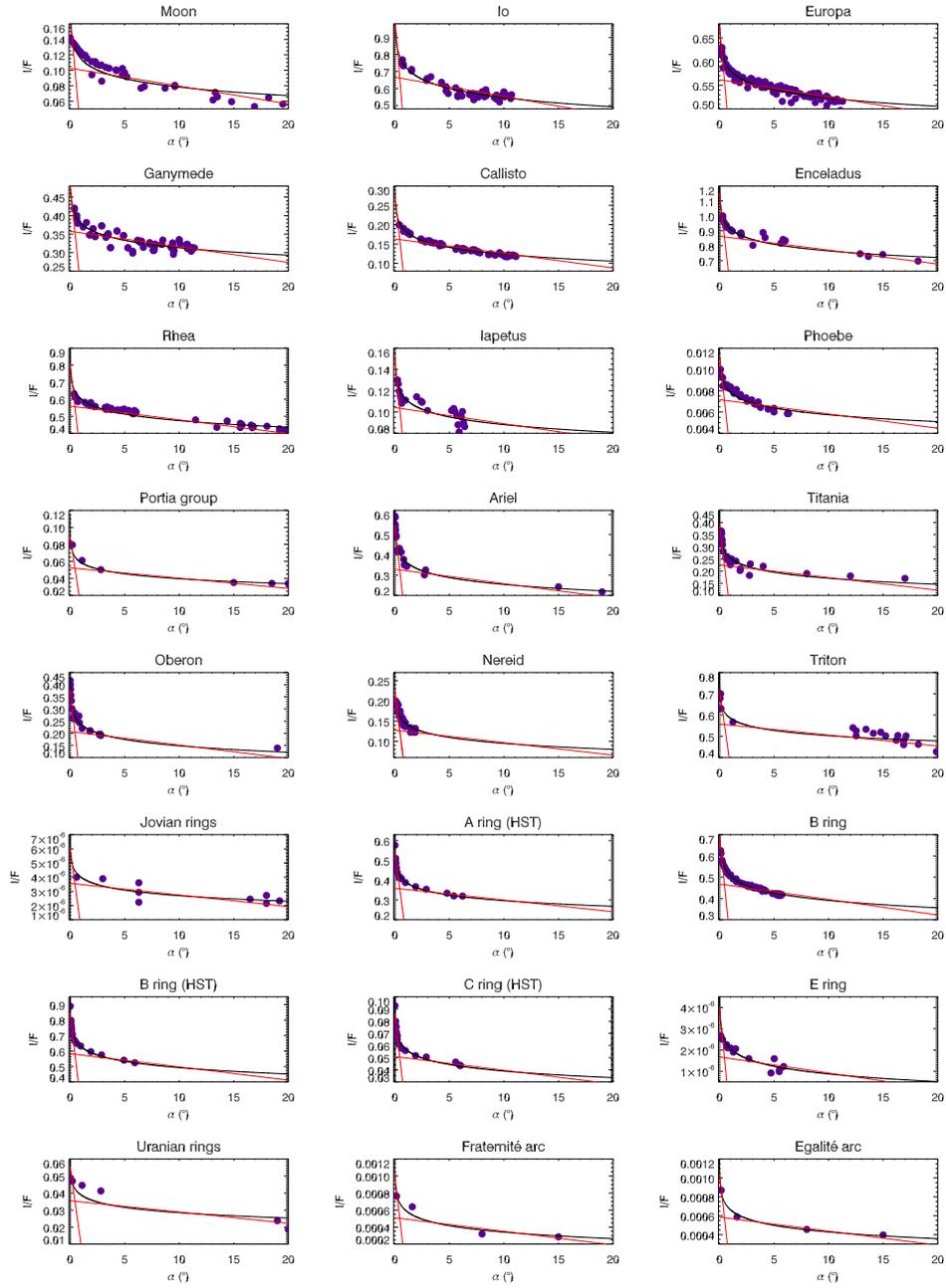}
\newline
\caption{\label{fig9} Phase curves of a selection of rings and satellites in
the Solar System. The solid lines correspond to the best fit obtained with
the linear-by-parts fit to the logarithmic model (in solid curves).}
\end{center}
\end{figure}

\newpage
\begin{figure}[!ht]
\includegraphics[width=14cm]{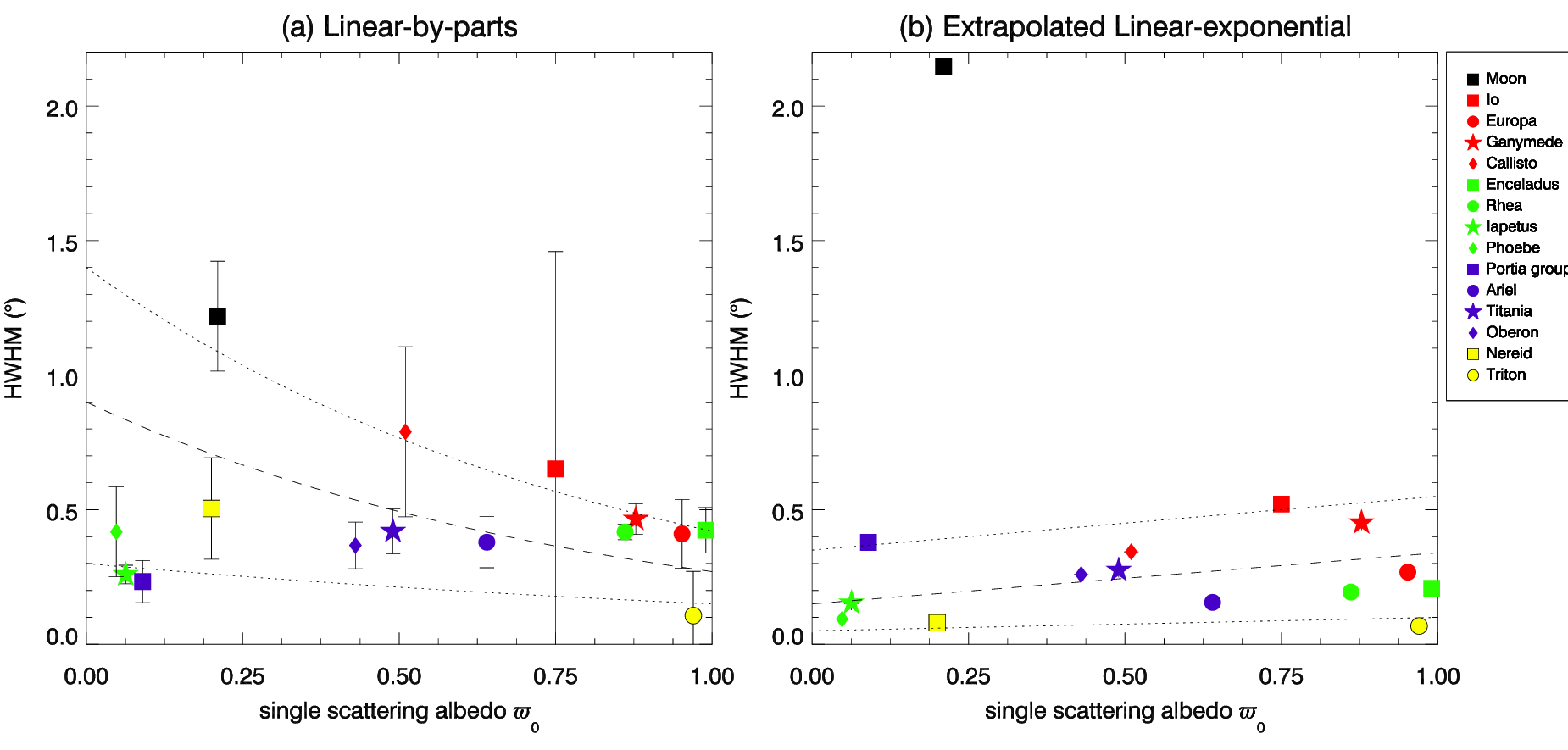}
\caption{\label{fig3} HWHM of the surge for satellites of Jupiter, Saturn, Uranus and Neptune derived with~: (a) the linear-by-parts model and (b) the linear-exponential model convolved with the size of the Sun. 
\newline
In (a) dashed line corresponds to a power-law fit to the data which is HWHM$\thicksim0.9\times0.3^{\varpi_{0}}$.
\newline
In (b) dashed line corresponds to a linear fit HWHM$\thicksim0.15+0.19\varpi_{0}$. 
\newline
Dotted lines are empirical functions to the boundaries of the data.}
\end{figure}

\newpage
\begin{figure}[!ht]
\includegraphics[width=14cm]{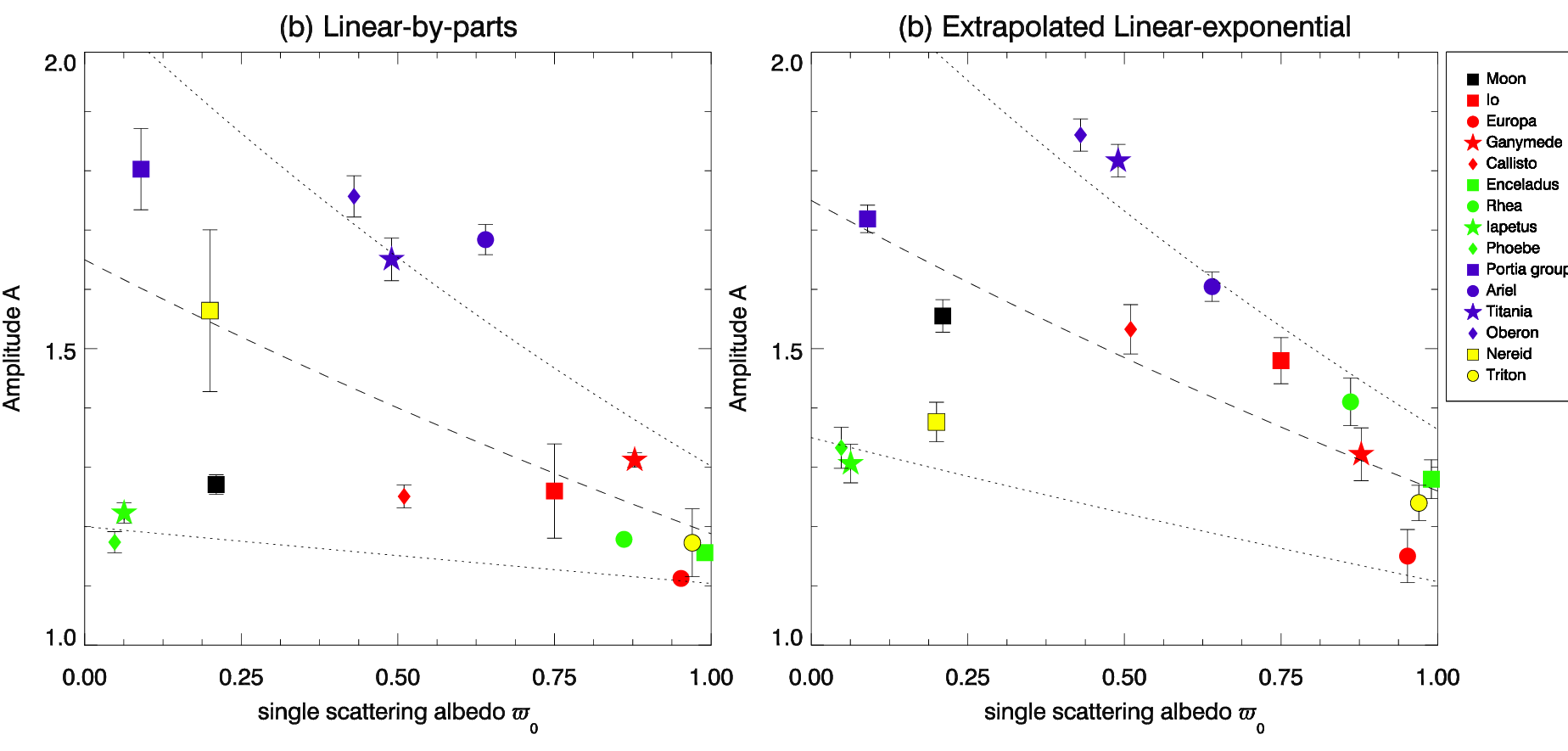}
\caption{\label{fig4} Amplitude of the surge derived with the linear-by-parts model for satellites of Jupiter, Saturn, Uranus and Neptune~: (a) the linear-by-parts model and (b) the linear-exponential model convolved with the size of the Sun. 
\newline
In (a) dashed line corresponds to a power fit to the data A$\thicksim1.65\times0.72^{\varpi_{0}}$
\newline
In (b) dashed line corresponds to the fit A$\thicksim1.75\times0.72^{\varpi_{0}}$. 
\newline
Dotted lines are empirical functions to the data boundaries.}
\end{figure}

\newpage
\begin{figure}[!ht]
\begin{center}
\includegraphics[width=14cm]{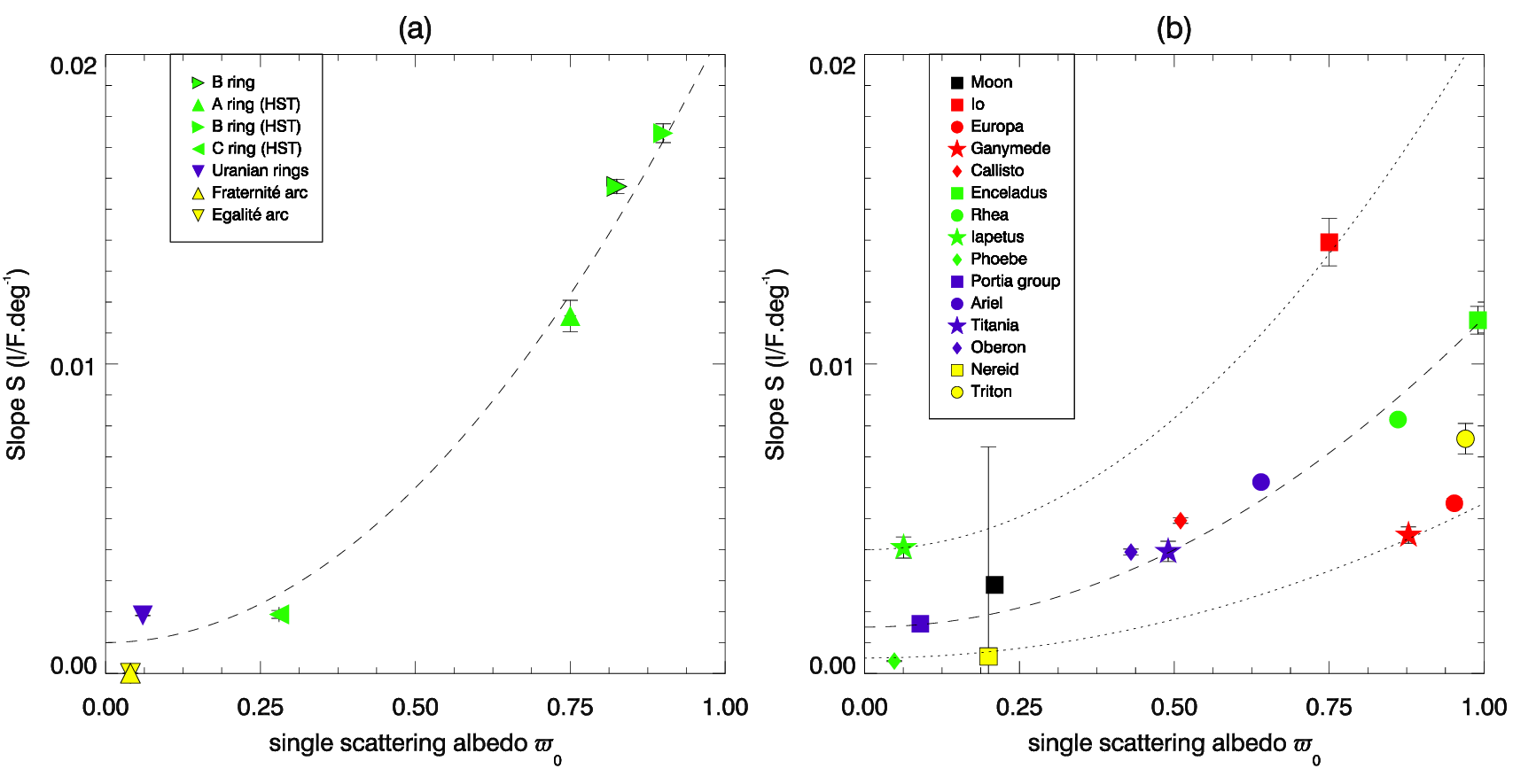}
\caption{\label{fig5} Morphological parameter~S derived with the linear-by-parts model for~: rings (a) and satellites (b) of Jupiter, Saturn, Uranus and Neptune. 
\newline
In~(a) dashed line corresponds  to a power-law fit to the data which is S$\thicksim0.001+0.02\varpi_{0}^{2}$ 
\newline
In (b) dashed line corresponds to a fit to the data which is S$\thicksim0.001+0.01\varpi_{0}^{2}$. 
\newline
Dotted lines are empirical functions to the data boundaries.}
\end{center}
\end{figure}

\newpage
\begin{figure}[!ht]
\includegraphics[width=14cm]{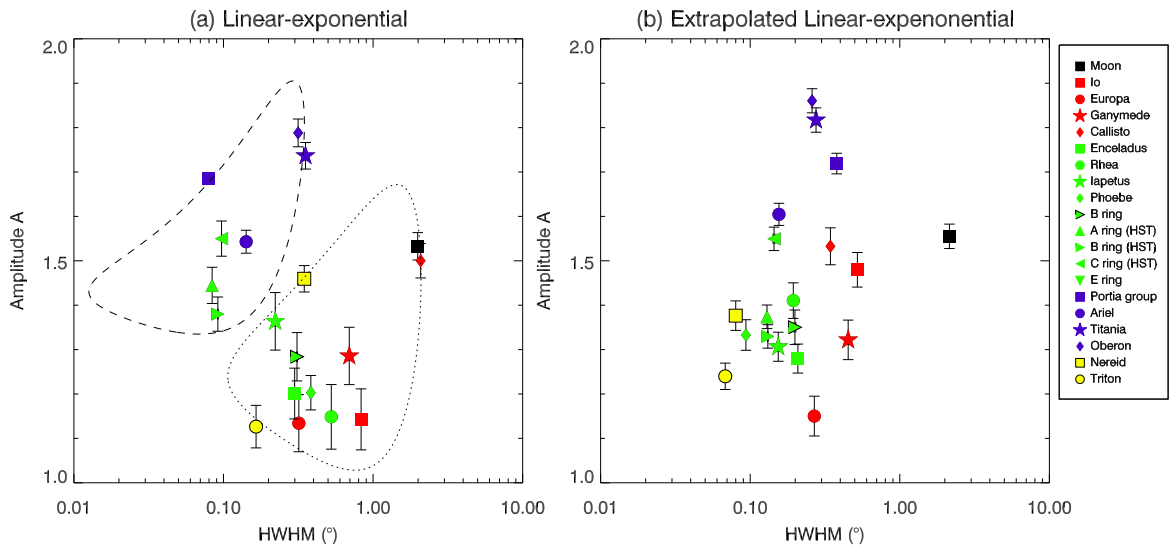}
\caption{\label{fig6} Cross comparison between the morphological parameters of the surge derived (a) with the linear-exponential model and (b) with the linear-exponential model convolved with the size of the Sun (see tables \,\ref{tab_solar_limb_dark_params_assbs} and \ref{tab_sunsize_assbs} of Appendix). 
\newline
Dashed and dotted ellipses are arbitrary delimitations of the data points (see text).}
\end{figure}

\newpage
\begin{figure}[!ht]
\includegraphics[width=14cm]{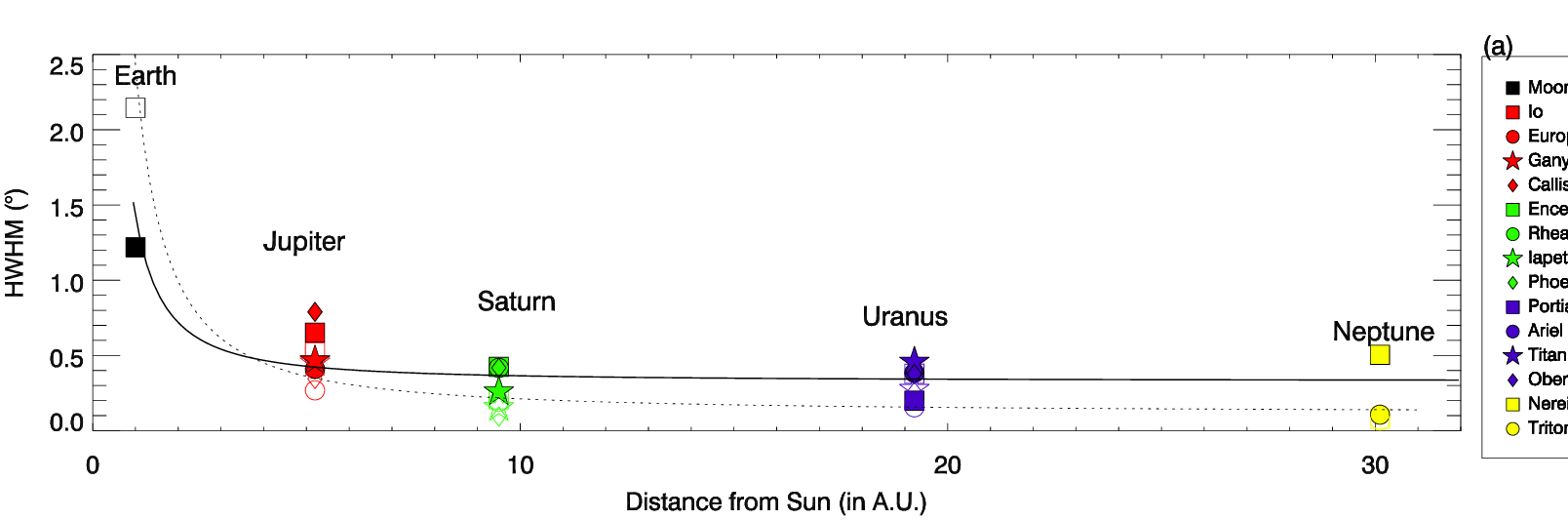}
\includegraphics[width=14cm]{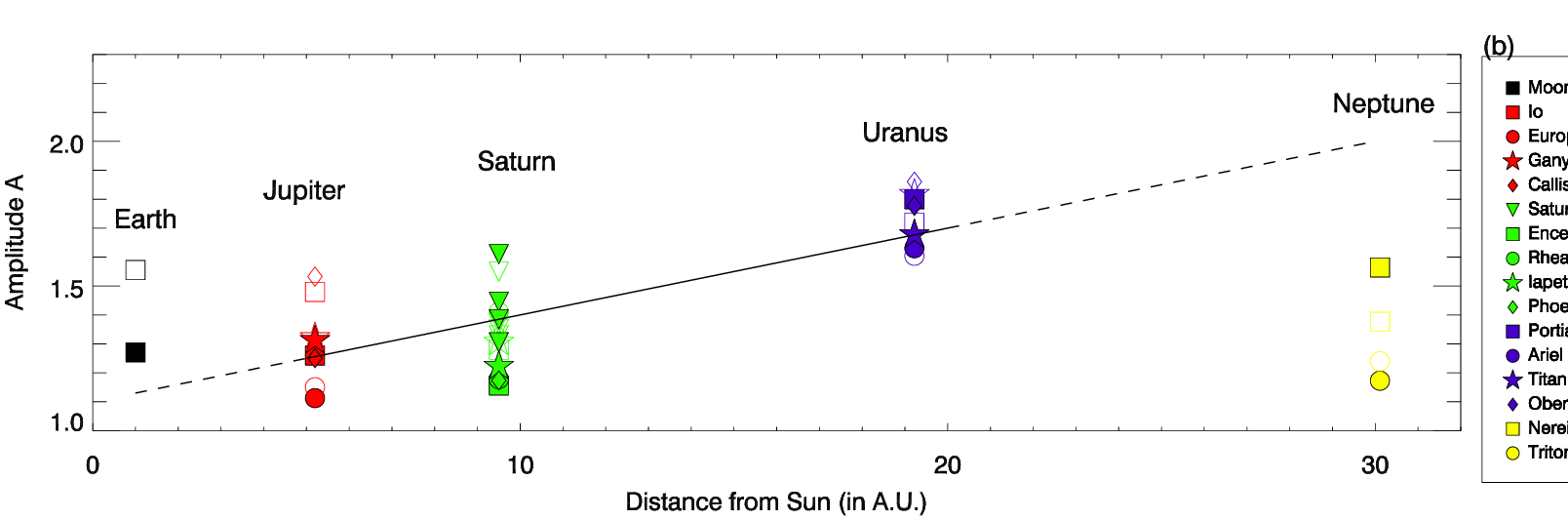}
\caption{\label{fig7} Variation of the morphological parameters HWHM and A derived with the linear-by-parts model (filled symbols and solid line) and the extrapolated linear-exponential model (empty symbols and dotted line) with respect to the distance from the Sun \citep[distances are taken from][]{2000ssd..book.....M}.}
\end{figure}

\newpage
\begin{figure}[!ht]
\begin{center}
\includegraphics[width=15cm]{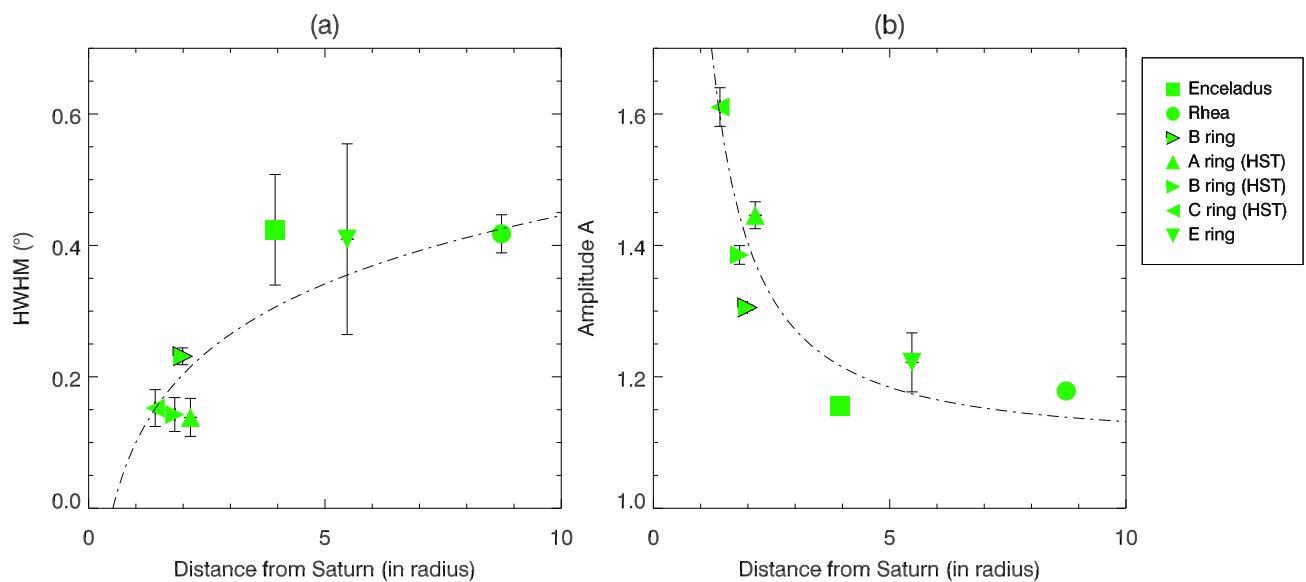}
\caption{\label{fig8} Variation of the morphological parameters: the amplitude A of the surge (a) and the half-width at half-maximum HWHM (b) derived with the linear-by-parts model with the distance from Saturn. 
\newline
\citep[Distances are taken from][]{2000ssd..book.....M}.}
\end{center}
\end{figure}

\newpage
\begin{figure}[!ht]
\begin{center}
\includegraphics[width=15cm]{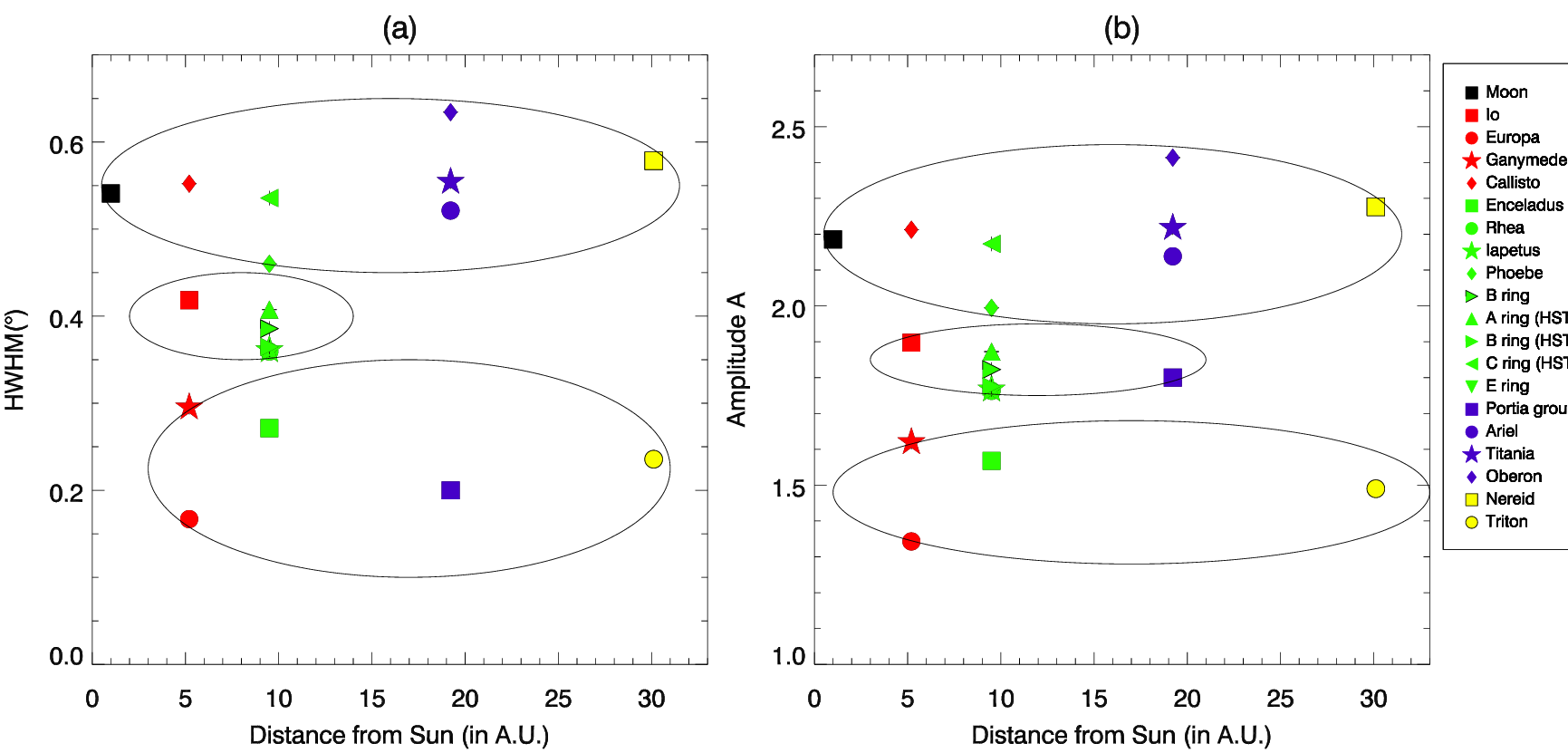}
\caption{\label{fig10}  \label{lastfig} Variation of the morphological parameters: the amplitude A of the surge (a) and the half-width at half-maximum HWHM (b) derived with the logarithmic model fitted by the linear-by-parts model with the distance from Saturn.
\newline
\citep[Distances are taken from][]{2000ssd..book.....M}. 
\newline
Solid ellipses in (a) and (b) are arbitrary delimitations of the data points (see text).}
\end{center}
\end{figure}

% for a selection of objects of the Solar System

%%%%%%%%%%%%%%%%%%%%%%%%%%%%%%%%%%%%%%%%%
%%%%%%%%%%%%%%%%%% tables
%%%%%%%%%%%%%%%%%%%%%%%%%%%%%%%%%%%%%%%%%%
\newpage

\begin{table}[!ht]
\begin{center}
\caption{\label{tab_phase_curves_assbs} References for 
opposition phase curves of the selection of Solar System rings and satellites.}
{\renewcommand{\arraystretch}{1}
%\small{
\begin{tabular}{llrl}
\hline
& \multirow{2}{1cm}{\centerline{Object}} & $\lambda$~~~ & \multirow{2}{3cm}{\centerline{~~References}} \\
&        &  (nm) &   \\
\hline
\hline
& Moon      & $\thicksim$570 & \citep{1969NASSP.201...38W,1933AnStr...2..203R}  \\
\hline
\multirow{6}{1em}{\rotatebox{90}{Jupiter}} 
& Main ring & $\thicksim$460 & \citep{2004Icar..172...59T}   \\
& Io        & $\thicksim$570 & \citep{1988Icar...75..450M}   \\
& Europa    & $\thicksim$500 & \citep{1992JGR....9714761T}   \\
& Ganymede  & $\thicksim$600 & \citep{1974Icar...23..399M,1975Icar...26..408M,1974AA....33..105B} \\
& Callisto  & $\thicksim$500 & \citep{1992JGR....9714761T}   \\
\hline
\multirow{10}{1em}{\rotatebox{90}{Saturn}}
& C ring    & 672 & \citep{2007PASP..119..623F}    \\
& B ring    & $\thicksim$650 & \citep{1965AJ.....70..704F}   \\ 
& B ring (HST)    & 672 & \citep{2007PASP..119..623F}   \\
& A ring    & 672 & \citep{2007PASP..119..623F}   \\
& E ring    & $\thicksim$650 & \citep{1983LPI....14..592P,1984prin.conf..111L,1991Icar...94..451S}    \\
& Enceladus & 439 & \citep{2005Icar..173...66V}  \\
& Rhea      & $\thicksim$500 & \citep{1995Icar..115..228D,1989Icar...82..336V}  \\
& Iapetus   & $\thicksim$600 & \citep{1974Icar...23..355F}  \\
& Phoebe    & $\thicksim$650 & \citep{2006Icar..184..181B} \\
\hline  
\multirow{5}{1em}{\rotatebox{90}{Uranus}}
& Rings     & $\thicksim$500 & \citep{2001Icar..151...51K}   \\
& Portia \textit{group} & $\thicksim$500 & \citep{2001Icar..151...51K}  \\
& Ariel     & $\thicksim$600 & \citep{1992AJ....104.1618B,2001Icar..151...51K}  \\
& Titania   & $\thicksim$600 & \citep{1992AJ....104.1618B,2001Icar..151...51K}  \\
& Oberon    & $\thicksim$600 & \citep{1992AJ....104.1618B,2001Icar..151...51K}   \\
\hline
\multirow{4}{1em}{\rotatebox{90}{Neptune}}
& Fraternit\'e   & $\thicksim$500 & \citep{2005Icar..174..263D,1994Icar..111..193F}   \\
& Egalit\'e     & $\thicksim$500 & \citep{2005Icar..174..263D,1994Icar..111..193F} \\
& Nereid      & $\thicksim$570 & \citep{2001Icar..151..112S}   \\
& Triton      & $\thicksim$400 & \citep{1991JGR....9619197B}   \\
\hline     
\end{tabular}}%}
\end{center}
\end{table}

\newpage

\begin{table}[!ht]
\begin{center}
\caption{\label{tab_ahwhms_assbs} Morphological parameters of opposition phase curves 
of Solar System objects. The unit of HWHM is the degree and the unit of the slope $S$ is I/F.deg$^{-1}$}
{\renewcommand{\arraystretch}{1}
%\small{
\begin{tabular}{llcccccc}
\hline
& \multirow{2}{1cm}{\centerline{Object}} &  \multicolumn{3}{c}{Linear-by-parts fit} &  \multicolumn{3}{c}{Linear-exponential fit} \\
\cline{3-8}
& &  A & HWHM & $S$ & A & HWHM & $S$  \\
\hline
\hline
& Moon & 1.27 & 1.21 & 0.0261 & 1.53 & 1.98 & 0.0017   \\
\hline
\multirow{5}{1em}{\rotatebox{90}{Jupiter}} 
& Main ring & 1.23 & 2.68 & 0.0170 & 1.34 & 2.03 & 4.82$\times10^{-8}$   \\
& Io & 1.25 & 0.65 & 0.0204 & 1.14 & 0.83 & 0.01500   \\
& Europa & 1.11 & 0.41 & 0.0095 & 1.13 & 0.31 & 0.0054   \\
& Ganymede & 1.31 & 0.46 & 0.0125 & 1.28 & 0.69 & 0.0040   \\
& Callisto & 1.25 & 0.78 & 0.0291 & 1.50 & 2.07 & 0.0024   \\
\hline
\multirow{10}{1em}{\rotatebox{90}{Saturn}}
& C ring & 1.61 & 0.15 & 0.0342 & 1.55 & 0.09 & 0.0030   \\
& B ring & 1.30 & 0.23 & 0.0312 & 1.28 & 0.30 & 0.0155   \\
& B ring (HST) & 1.38 & 0.14 & 0.0278 & 1.37 & 0.09 & 0.0238   \\
& A ring & 1.44 & 0.13 & 0.0297 & 1.44 & 0.08 & 0.0159   \\
& E ring & 1.22 & 0.40 & 0.0903 & 1.51 & 0.86 & 1.19$\times10^{-7}$   \\
& Enceladus & 1.15 & 0.42 & 0.0127 & 1.20 & 0.29 & 0.0113   \\
& Rhea & 1.17 & 0.41 & 0.0142 & 1.14 & 0.52 & 0.0080   \\
& Iapetus & 1.22 & 0.25 & 0.0348 & 1.36 & 0.22 & 0.0030   \\
& Phoebe & 1.17 & 0.41 & 0.0485 & 1.20 & 0.38 & 0.0003   \\
\hline  
\multirow{5}{1em}{\rotatebox{90}{Uranus}}
& Rings & 1.07 & 0.21 & 0.0274 & 1.08 & 0.45 & 0.0012   \\
& Portia \textit{group} & 1.80 & 0.20 & 0.0015 & 1.48 & 0.13 & 0.0015   \\
& Ariel & 1.63 & 0.38 & 0.0178 & 1.60 & 0.18 & 0.0074   \\
& Titania & 1.67 & 0.45 & 0.0126 & 1.82 & 0.27 & 0.0032   \\
& Oberon & 1.77 & 0.37 & 0.0174 & 1.81 & 0.31 & 0.0037   \\
\hline
\multirow{4}{1em}{\rotatebox{90}{Neptune}}
& Fraternit\'e & 2.16 & 2.57 & 0.0136 & 1.86 & 1.50 & 1.10$\times10^{-5}$   \\
& Egalit\'e & 1.72 & 1.01 & 0.0160 & 1.77 & 0.67 & 8.34$\times10^{-6}$   \\
& Nereid & 1.56 & 0.50 & 0.0042 & 1.45 & 0.34 & 0.0099   \\
& Triton & 1.17 & 0.10 & 0.0124 & 1.25 & 0.27 & 0.0049   \\
\hline     
\end{tabular}}%}
\end{center}
\end{table}

\newpage
\begin{table}[!ht]
%\begin{center}
\caption{\label{tab_all_params_assbs} Direct output parameters of morphological models using opposition phase curves and ``ideal'' opposition phase curves of Solar System objects. The unit of $w$ is the degree and the unit of the slope $I_{s}$ is I/F.deg$^{-1}$}
{\renewcommand{\arraystretch}{1}
\tiny{
%\begin{sideways} 
\begin{tabular}{llcccccccccc}
\hline
& \multirow{2}{1cm}{\centerline{Object}} & \multicolumn{2}{c}{Logarithmic fit} & \multicolumn{4}{c}{Linear-exponential fit} & \multicolumn{4}{c}{Extrapolated linear-exponential fit} \\
\cline{3-12}
&  & $a_{0}$ & $a_{1}$ & $I_{p}$ & $I_{b}$ & $w$ &  $I_{s}$ & $I_{p}$ & $I_{b}$ & $w$ &  $I_{s}$ \\
\hline
\hline
& Moon & 0.114 & -0.015 & 0.048 & 0.090 & 1.433 & 0.0017 & 0.049 & 0.088 & 1.548 & 0.0016  \\
\hline
\multirow{5}{1em}{\rotatebox{90}{Jupiter}} 
& Main ring & 3.9$\times10^{-6}$ & -5.6$\times10^{-7}$ & 1.1$\times10^{-6}$ & 3.2$\times10^{-6}$ & 1.470 & 4.8$\times10^{-8}$ & 2.4$\times10^{-6}$ & 3.2$\times10^{-6}$ & 0.905 & 4.6$\times10^{-8}$  \\
& Io & 0.726 & -0.077 & 0.100 & 0.699 & 0.600 & 0.0150 & 0.313 & 0.652 & 0.375 & 0.0107  \\
& Europa & 0.580 & -0.025 & 0.076 & 0.572 & 0.230 & 0.0054 & 0.086 & 0.573 & 0.193 & 0.0055  \\
& Ganymede & 0.379 & -0.028 & 0.100 & 0.349 & 0.500 & 0.0040 & 0.114 & 0.355 & 0.325 & 0.0041  \\
& Callisto & 0.182 & -0.025 & 0.070 & 0.140 & 1.500 & 0.0024 & 0.091 & 0.171 & 0.248 & 0.0051  \\
\hline
\multirow{10}{1em}{\rotatebox{90}{Saturn}}
& C ring & 0.056 & -0.007 & 0.033 & 0.059 & 0.070 & 0.0030 & 0.031 & 0.056 & 0.104 & 0.0021  \\
& B ring & 0.506 & -0.050 & 0.142 & 0.502 & 0.223 & 0.0155 & 0.179 & 0.511 & 0.143 & 0.0174  \\
& B ring (HST) & 0.628 & -0.058 & 0.249 & 0.656 & 0.066 & 0.0238 & 0.213 & 0.647 & 0.094 & 0.0216  \\
& A ring & 0.390 & -0.046 & 0.182 & 0.409 & 0.060 & 0.0159 & 0.150 & 0.401 & 0.093 & 0.0143  \\
& E ring & 2.1$\times10^{-6}$ & -5.2$\times10^{-7}$ & 9.3$\times10^{-7}$ & 1.8$\times10^{-6}$ & 0.623 & 1.2$\times10^{-7}$ & 4.8$\times10^{-7}$ & 1.2$\times10^{-6}$ & 1.085 & 3.8$\times10^{-7}$  \\
& Enceladus & 0.912 & -0.063 & 0.180 & 0.895 & 0.215 & 0.0113 & 0.251 & 0.899 & 0.150 & 0.0116  \\
& Rhea & 0.602 & -0.055 & 0.085 & 0.573 & 0.379 & 0.0080 & 0.236 & 0.576 & 0.139 & 0.0082  \\
& Iapetus & 0.112 & -0.010 & 0.039 & 0.110 & 0.160 & 0.0030 & 0.035 & 0.116 & 0.111 & 0.0039  \\
& Phoebe & 0.007 & -0.0001 & 0.001 & 0.008 & 0.276 & 0.0003 & 0.002 & 0.008 & 0.067 & 0.0004  \\
\hline  
\multirow{5}{1em}{\rotatebox{90}{Uranus}}
& Rings & 0.038 & -0.004 & 0.003 & 0.044 & 0.326 & 0.0012 & 0.014 & 0.045 & 0.017 & 0.0012  \\
& Portia \textit{group} & 0.058 & -0.006 & 0.026 & 0.055 & 0.100 & 0.0015 & 0.038 & 0.054 & 0.273 & 0.0010  \\
& Ariel & 0.366 & -0.044 & 0.213 & 0.355 & 0.129 & 0.0074 & 0.217 & 0.359 & 0.112 & 0.0077  \\
& Titania & 0.252 & -0.034 & 0.181 & 0.221 & 0.198 & 0.0032 & 0.181 & 0.221 & 0.198 & 0.0032  \\
& Oberon & 0.236 & -0.034 & 0.168 & 0.208 & 0.229 & 0.0037 & 0.181 & 0.211 & 0.186 & 0.0040  \\
\hline
\multirow{4}{1em}{\rotatebox{90}{Neptune}}
& Fraternit\'e & 0.0005 & -0.0001 & 0.00037 & 0.00043 & 1.082 & 1.1$\times10^{-5}$ & 0.00086 & 0.00022 & 1.1$\times10^{-5}$ & 5.6$\times10^{-6}$  \\
& Egalit\'e & 0.0001 & -0.001 & 0.00040 & 0.00052 & 0.489 & 8.3$\times10^{-6}$ & 0.00090 & 0.00021 & 1.106 & 1.5$\times10^{-5}$  \\
& Nereid & 0.144 & -0.021 & 0.064 & 0.140 & 0.250 & 0.0099 & 0.065 & 0.175 & 0.057 & 0.0301  \\
& Triton & 0.584 & -0.035 & 0.140 & 0.560 & 0.200 & 0.0049 & 0.146 & 0.609 & 0.049 & 0.0075  \\
\hline     
\end{tabular}}}
%\end{sideways}}}
%\end{center}
\end{table}

\newpage
\begin{table}[!ht]
\begin{center}
\caption{\label{tab_albedo_assbs} References for the single scattering albedo of Solar System objects.}
{\renewcommand{\arraystretch}{1}
%\small{
\begin{tabular}{lllrl}
\hline
& \multirow{2}{1cm}{\centerline{Object}} & \multirow{2}{0.7cm}{\centerline{$\varpi_{0}$}} &  $\lambda$~~~ & \multirow{2}{3cm}{\centerline{~~References}} \\
&       &  &  (nm) &   \\
\hline
\hline
& Moon         & 0,21 & $\thicksim$500 &  \citep{1997Icar..128....2H}  \\
\hline
\multirow{5}{1em}{\rotatebox{90}{Jupiter}} 
& Io           & 0,75 & 590 & \citep{1988Icar...75..450M}   \\
& Europa       & 0,96 & 550 & \citep{1997Icar..128...49D}   \\
& Ganymede     & 0,87 & 470 & \citep{1997Icar..128...49D}   \\ 
& Callisto     & 0,53 & 470 & \citep{1997Icar..128...49D}   \\
\hline
\multirow{9}{1em}{\rotatebox{90}{Saturn}}
& C ring       & 0,16 & 672 & \citep{2007PASP..119..623F}   \\
& B ring       & 0,83 & 672 & \citep{2002Icar..158..224P}   \\ 
& B ring (HST) & 0,85 & 672 & \citep{2007PASP..119..623F}   \\
& A ring       & 0,79 & 672 & \citep{2007PASP..119..623F}   \\
& Enceladus    & 0,99 & 480 & \citep{1991BAAS...23.1168V}   \\
& Rhea         & 0,86 & 480 & \citep{1989Icar...82..336V}   \\
& Iapetus      & 0,16 & 480 & \citep{1984Icar...59..392B}   \\
& Phoebe       & 0,06 & 480 & \citep{1999Icar..138..249S}   \\
\hline  
\multirow{5}{1em}{\rotatebox{90}{Uranus}}
& Rings        & 0,06 & $\thicksim$500 & \citep{2001Icar..151...51K}   \\
& Portia \textit{group} & 0,09 & $\thicksim$500 & \citep{2001Icar..151...51K}  \\
& Ariel        & 0,64 & $\thicksim$500 & \citep{2001Icar..151...51K}  \\
& Titania      & 0,48 & $\thicksim$475 & \citep{1987JGR....9214895V}   \\
& Oberon       & 0,43 & $\thicksim$500 & \citep{2001Icar..151...51K}   \\
\hline
\multirow{4}{1em}{\rotatebox{90}{Neptune}}
& Fraternit\'e  & 0,02 & 480 & \citep{1994Icar..111..193F}  \\
& Egalit\'e    & 0,02 & 480 & \citep{1994Icar..111..193F}  \\
& Nereid     & 0,21 & $\thicksim$500 & \citep{1991JGR....9619253T}  \\
& Triton     & 0,97 & 500 & \citep{1992Icar...99...82L}   \\
\hline     
\end{tabular}}%}
\end{center}
\end{table}

\newpage
\begin{table}[!ht]
\begin{center}
\caption{\label{tab_ahwhms_ideal_assbs} \label{lasttable} Morphological parameters of ``ideal'' opposition phase curves 
of Solar System objects. The unit of HWHM is the degree and the unit of the slope $S$ is I/F.deg$^{-1}$}
{\renewcommand{\arraystretch}{1}
%\small{
\begin{tabular}{llcccccc}
\hline
& \multirow{2}{1cm}{\centerline{Object}} &  \multicolumn{3}{c}{Linear-by-parts and log fits} &  \multicolumn{3}{c}{Extrapolated linear-exponential fit} \\
\cline{3-8}
& &  A & HWHM & $S$ & A & HWHM & $S$  \\
\hline
\hline
& Moon & 2.18 & 0.54 & 0.0224 & 1.55 & 2.14 & 0.0016   \\
\hline
\multirow{5}{1em}{\rotatebox{90}{Jupiter}} 
& Main ring & 2.21 & 0.55 & 0.0230 & 1.76 & 1.25 & 4.55$\times10^{-8}$   \\
& Io & 1.89 & 0.41 & 0.0169 & 1.47 & 0.52 & 0.0107   \\
& Europa & 1.34 & 0.16 & 0.0064 & 1.15 & 0.26 & 0.0055   \\
& Ganymede & 1.62 & 0.29 & 0.0117 & 1.32 & 0.45 & 0.0041   \\
& Callisto & 2.21 & 0.55 & 0.0229 & 1.53 & 0.34 & 0.0051   \\
\hline
\multirow{10}{1em}{\rotatebox{90}{Saturn}}
& C ring & 2.17 & 0.53 & 0.0221 & 1.54 & 0.14 & 0.0021   \\
& B ring & 1.82 & 0.38 & 0.0155 & 1.35 & 0.19 & 0.0174   \\
& B ring (HST) & 1.77 & 0.36 & 0.0146 & 1.32 & 0.13 & 0.0216   \\
& A ring & 1.87 & 0.40 & 0.0165 & 1.37 & 0.12 & 0.0143   \\
& E ring & 3.38 & 0.99 & 0.0450 & 2.80 & 1.50 & 3.76$\times10^{-7}$   \\
& Enceladus & 1.56 & 0.27 & 0.0107 & 1.27 & 0.20 & 0.0116   \\
& Rhea & 1.76 & 0.35 & 0.0144 & 1.41 & 0.19 & 0.0082   \\
& Iapetus & 1.76 & 0.36 & 0.0145 & 1.30 & 0.15 & 0.0039   \\
& Phoebe & 1.99 & 0.45 & 0.0187 & 1.33 & 0.09 & 0.0004   \\
\hline  
\multirow{5}{1em}{\rotatebox{90}{Uranus}}
& Rings & 1.98 & 0.45 & 0.0186 & 1.31 & 0.02 & 0.0012   \\
& Portia \textit{group} & 1.80 & 0.20 & 0.0015 & 1.71 & 0.37 & 0.0010   \\
& Ariel & 2.13 & 0.52 & 0.0215 & 1.60 & 0.15 & 0.0077   \\
& Titania & 2.21 & 0.55 & 0.0230 & 1.81 & 0.27 & 0.0032   \\
& Oberon & 2.41 & 0.63 & 0.0267 & 1.86 & 0.25 & 0.0040   \\
\hline
\multirow{4}{1em}{\rotatebox{90}{Neptune}}
& Fraternit\'e & 2.72 & 0.75 & 0.0314 & 4.99 & 1.53 & 5.61$\times10^{-6}$    \\
& Egalit\'e & 2.39 & 0.62 & 0.0253 & 5.34 & 1.53 & 1.53$\times10^{-5}$   \\
& Nereid & 2.27 & 0.57 & 0.0241 & 1.37 & 0.08 & 0.0301   \\
& Triton & 1.49 & 0.23 & 0.0092 & 1.23 & 0.06 & 0.0075   \\
\hline     
\end{tabular}}%}
\end{center}
\end{table}

\newpage
\begin{table}[!ht]
\textbf{APPENDIX : \\
REFINEMENTS TO THE SUN'S ANGULAR SIZE EFFECT}
To compute the distance from a Solar System object to the Sun at any given date, it is not appropriate to use the semi-major axis. We 
use a series of equations which take into consideration the distance between the Sun and the planet at the approximate date.
The heliocentric radius $r_{\textrm{p}}$, the distance from the focus of the ellipse (i.e. the Sun) to the planet, is given by:
\begin{equation}
r_{\textrm{p}}=a(1-e\cos E)
\end{equation}
where $E$ is the eccentric anomaly, $a$ and $e$ are two of the seven orbital elements which define an ellipse in space~: $a$ is the mean distance, or the value of the semi-major axis of the orbit (average Sun to planet distance); $e$ is the eccentricity of the ellipse which describes the orbit (dimensionless); $i$ is the inclination (in degrees), or angle between the plane of the ecliptic (the plane of the Earth's orbit about the Sun) and the plane of the planets orbit; $\Omega$ is the longitude of ascending node (in degrees), or the position in the orbit where the elliptical path of the planet passes through the plane of the ecliptic, from below the plane to above the plane; $\tilde{\omega}$ is the longitude of perihelion (in degrees), or the position in the orbit where the planet is closest to the Sun; $\lambda$ is the mean longitude (in degrees), the position of the planet in the orbit; and $M$ is mean anomaly (in degrees).
The mean anomaly gives the planet's angular position for a circular orbit with radius equal to the semi major axis. It is computed directly from the elements using:
\begin{equation}
M=\lambda-\tilde{\omega}
\end{equation}
Kepler's second law states that the radius vector of a planet sweeps out equal areas in equal times. The planet must speed up and slow down in its orbit. The true anomaly~$f$ gives the planet's actual angular position in its orbit. It is the angle (at the Sun) between perihelion of the orbit and the current location of the planet. 
To obtain its value, first we compute the eccentric anomaly, $E$, from $M$ and the eccentricity $e$ by using the ``Kepler equation'':
\end{table}

\newpage
\begin{table}[!ht]

\begin{equation}
M=E-e\sin E
\end{equation}
An expansion to order $e^3$ of the solution to the ``Kepler equation'' is:
\begin{equation}
E=M+ (e-\frac{e^{3}}{8})\sin M +\frac{1}{2}e^{2}\sin 2M + \frac{3}{8}e^{3}\sin 3M
\end{equation}
%However another way to compute $M$ could be to use as a first approximation:
%\begin{equation}
%E = M + e\sin M (1 + e\cos M)
%\end{equation}

%and then iterate using:
%\begin{eqnarray}
%E' &=& E \\
%E  &=& E' -  \frac{E' - e\sin E' - M}{1 - e\cos E'}
%\end{eqnarray}
%until the magnitude of $E - E'$ is sufficiently close to zero. Finally, converting the eccentric anomaly to the true anomaly is possible by using:
%\begin{equation}
%f = 2\arctan\left[\sqrt{\frac{(1 + e)}{(1 - e)}}\cdot \tan\left(\frac{1}{2}E\right)\right]
%\end{equation}
%and ensure that the result is in the range 0-360\textsuperscript{o}. In this way, the heliocentric radius is given by a formula based on the geometry of an ellipse:
%\begin{equation}
%r_{p} =\frac{a\cdot(1 - e^{2})}{1 + e\cos f}
%\end{equation}
         
Now, to find the orbital elements of a planet at a specific date, we use:    
\begin{eqnarray}
a &=& a_{0} + \dot{a}t \\
e &=& e_{0} + \dot{e}t \\
i &=& i_{0} + \left(\dot{a}/3600\right)t \\
\tilde{\omega} &=& \tilde{\omega}_{0} + \left(\dot{\tilde{\omega}}/3600\right)t \\
\Omega &=& \Omega_{0} + \left(\dot{\Omega}/3600\right)t \\
\lambda &=& \lambda_{0} + \left(\dot{\lambda}/3600+360~N_{r}\right)t 
\end{eqnarray}       
where $t$ is the observation time (Table~Appendix\ref{tab_sunsize_assbs}) converted in Julian centuries since JD~2451545.0, the zero index quantities are the orbital elements at the epoch of J2000 (JD~2451545.0) and the dot quantities are the change per julian century of the orbital elements \citep[values are taken in][]{2000ssd..book.....M}.  
When we have the heliocentric radius at the given observation time, we can compute the solar angular size $\alpha_{\odot}$ using $r_{p}$ and $r_{\odot}$, the radius of the Sun:
\begin{equation}
\alpha_{\odot} =\arcsin\frac{r_{\odot}}{r_{p}}
\end{equation}
The limb darkening function of \citet{1961MmRAS..63...89P} has been used in the past to be convolved with a theoretical opposition effect function \citep{1974IAUS...65..441K}. Its formula is: 
\begin{equation}
W(\mu')= a_{\lambda} + b_{\lambda}\mu' + c_{\lambda}\left[1-\mu'\cdot \log\left(1 + \frac{1}{\mu'}\right)\right]
\end{equation}
where $\mu'=\cos\theta'$ and $\theta'$ varies from 0 to the Sun's angular radius $\alpha_{\odot}$. $a_{\lambda}$, $b_{\lambda}$ and $c_{\lambda}$ are coefficients that depend on the wavelength (values are given in table~Appendix\ref{tab_solar_limb_dark_params_assbs}).
We perform a normalized convolution of the limb darkening function to the linear-exponential function~$P(\alpha)$ by doing:
\begin{equation}
P'(\alpha)= \frac{\int_{0}^{\alpha_{\odot}} P(\alpha)\cdot W(\cos\theta') d\theta'}{\int_{0}^{\alpha_{\odot}} W(\cos\theta') d\theta'}
\end{equation}

\end{table}

\newpage
%\multirow{2}{3.5cm}{\centerline{}}
\begin{table}[!ht]
\begin{center}
\caption{\label{tab_sunsize_assbs} References for 
the observational parameters needed to compute the angular size of the Sun $\alpha_{\odot}$. Bold text corresponds to missing data replaced arbitrarily, we assumed an hourly time of 18:00:00.0}
{\renewcommand{\arraystretch}{1}
%\footnotesize{
\begin{tabular}{llcll}
\hline
& \multirow{2}{1cm}{\centerline{Object}} & Solar Angular Size & Observation Time & \multirow{2}{3cm}{\centerline{~~References}} \\
&     &  $\alpha_{\odot}$(\textsuperscript{o}) & UT &     \\
\hline
\hline
& Moon      & 0.263 & 1968 December 24 & \citep{1969NASSP.201...38W}  \\
\hline
\multirow{6}{1em}{\rotatebox{90}{Jupiter}} 
& Main ring   & 0.0506 & 2000 December 13 & \citep{2004Icar..172...59T}   \\
& Io        & 0.0498 & 1977 December 15 & \citep{1980AJ.....85..961L}   \\
& Europa    & 0.0515 & 1976 \textbf{January 1} & \citep{1992JGR....9714761T}   \\
& Ganymede  & 0.0499 & 1971 may \textbf{1} & \citep{1974AA....33..105B} \\
& Callisto  & 0.0515 & 1976 \textbf{January 1} & \citep{1992JGR....9714761T}   \\
\hline
\multirow{10}{1em}{\rotatebox{90}{Saturn}}
& C ring    & 0.0295 & 2005 January 13 & \citep{2007PASP..119..623F}    \\
& B ring    & 0.0266 & 1959 June 26 & \citep{1965AJ.....70..704F}   \\ 
& B ring (HST) & 0.0295 & 2005 January 13 & \citep{2007PASP..119..623F}   \\
& A ring    & 0.0295 & 2005 January 13 & \citep{2007PASP..119..623F}   \\
& E ring    & 0.0271 & 1980 \textbf{January 1} & \citep{1984prin.conf..111L}    \\
& Enceladus & 0.0287 & 1997 October 10 & \citep{2005Icar..173...66V}  \\
& Rhea      & 0.0268 & 1976 January 13 & \citep{1980AJ.....85..961L}  \\
& Iapetus   & 0.0266 & 1972 December \textbf{1} & \citep{1974Icar...23..355F}  \\
& Phoebe    & 0.0295 & 2005 January 13 & \citep{2006Icar..184..181B} \\
\hline  
\multirow{5}{1em}{\rotatebox{90}{Uranus}}
& Rings     & 0.0138 & 1997 July 29 & \citep{2001Icar..151...51K}   \\
& Portia \textit{group} & 0.0138 & 1997 July 29 & \citep{2001Icar..151...51K}  \\
& Ariel     & 0.0138 & 1997 July 29 & \citep{2001Icar..151...51K}  \\
& Titania   & 0.0138 & 1997 July 29 & \citep{2001Icar..151...51K}  \\
& Oberon    & 0.0138 & 1997 July 29 & \citep{2001Icar..151...51K}   \\
\hline
\multirow{4}{1em}{\rotatebox{90}{Neptune}}
& Fraternit\'e   & 0.00885 & 2002 July 27 & \citep{2005Icar..174..263D}   \\
& Egalit\'e     & 0.00885 & 2002 July 27 & \citep{2005Icar..174..263D} \\
& Nereid      & 0.00889 & 1998 June 20 & \citep{2001Icar..151..112S}   \\
& Triton      & 0.00885 & \textbf{1988 June 20} & \citep{1991JGR....9619197B}   \\
\hline     
\end{tabular}}%}
\end{center}
\end{table}

\newpage
\begin{table}[!ht]
\begin{center}
\caption{\label{tab_solar_limb_dark_params_assbs} Parameters of the limb darkening function of \citet{1961MmRAS..63...89P} at similar wavelengths to the observations.}
{\renewcommand{\arraystretch}{1}
%\footnotesize{
\begin{tabular}{llrrccc}%rcccccc}
\hline
& \multirow{3}{1cm}{\centerline{Object}} &  & \multicolumn{4}{c}{\citet{1961MmRAS..63...89P}} \\ %&  \multicolumn{7}{c}{\citet{1994SoPh..153...91N}} \\ 
\cline{4-7}
& & $\lambda_{\textrm{obs}}$ & $\lambda$ & $a_{\lambda}$ & $b_{\lambda}$ & $c_{\lambda}$ \\ %& $\lambda$ & $A_{0}$ & $A_{1}$ & $A_{2}$ & $A_{3}$ & $A_{4}$ & $A_{5}$ \\
&        &  (nm) & (nm)  & & & \\ %& (nm) & & & & & &  \\
\hline
\hline
& Moon      & $\thicksim$570 & 560 & 0.75079 & 0.41593 & -0.54334 \\ %& 559.9 & 0.26892 & 1.34319 & -1.58427 & 1.91271 & -1.31530 & 0.37295 \\
\hline
\multirow{6}{1em}{\rotatebox{90}{Jupiter}} 
& Main ring    & $\thicksim$460 & 460 & 0.58274 & 0.56078 & -0.46772 \\ %& 457.3 & 0.16604 & 1.38544 & -1.52275 & 2.00232 & -1.45969 & 0.42864 \\
& Io        & $\thicksim$570 & 560 & 0.75079 & 0.41593 & -0.54334 \\ %& 559.9 & 0.26892 & 1.34319 & -1.58427 & 1.91271 & -1.31530 & 0.37295 \\
& Europa    & $\thicksim$500 & 500 & 0.68897 & 0.47873 & -0.54651 \\ %& 492.9 & 0.20924 & 1.30798 & -1.20411 & 1.21505 & -0.67196 & 0.14381 \\
& Ganymede  & $\thicksim$600 & 600 & 0.78074 & 0.38427 & -0.53777 \\ %& 610.9 & 0.30854 & 1.36620 & -1.83572 & 2.33221 & -1.63082 & 0.45959 \\
& Callisto  & $\thicksim$500 & 500 & 0.68897 & 0.47873 & -0.54651 \\ %& 492.9 & 0.20924 & 1.30798 & -1.20411 & 1.21505 & -0.67196 & 0.14381 \\
\hline
\multirow{10}{1em}{\rotatebox{90}{Saturn}}
& C ring    & 672 & 660 & 0.83717 & 0.33283 & -0.55400 \\ %& 669.4 & 0.34685 & 1.37539 & -2.04425 & 2.70493 & -1.94290 & 0.55999 \\
& B ring    & $\thicksim$650 & 640 & 0.81999 & 0.34918 & -0.55132 \\ % & 640.9 & 0.33644 & 1.30590 & -1.79238 & 2.45040 & -1.89979 & 0.59943 \\ 
& B ring (HST)   & 672 & 660 & 0.83717 & 0.33283 & -0.55400 \\ %& 669.4 & 0.34685 & 1.37539 & -2.04425 & 2.70493 & -1.94290 & 0.55999 \\
& A ring    & 672 & 660 & 0.83717 & 0.33283 & -0.55400 \\ %& 669.4 & 0.34685 & 1.37539 & -2.04425 & 2.70493 & -1.94290 & 0.55999 \\
& E ring    & $\thicksim$650 & 640 & 0.81999 & 0.34918 & -0.55132 \\ %& 640.9 & 0.33644 & 1.30590 & -1.79238 & 2.45040 & -1.89979 & 0.59943 \\
& Enceladus & 439 & 440 & 0.49375 & 0.62584 & -0.38974 \\ %& 443.9 & 0.16220 & 1.24893 & -0.92165 & 0.89978 & -0.50148 & 0.11220 \\
& Rhea      & $\thicksim$500 & 500 & 0.68897 & 0.47873 & -0.54651 \\ %& 492.9 & 0.20924 & 1.30798 & -1.20411 & 1.21505 & -0.67196 & 0.14381 \\
& Iapetus   & $\thicksim$600 & 600 & 0.78074 & 0.38427 & -0.53777 \\ %& 610.9 & 0.30854 & 1.36620 & -1.83572 & 2.33221 & -1.63082 & 0.45959 \\
& Phoebe    & $\thicksim$650 & 640 & 0.81999 & 0.34918 & -0.55132 \\ %& 640.9 & 0.33644 & 1.30590 & -1.79238 & 2.45040 & -1.89979 & 0.59943 \\
\hline  
\multirow{5}{1em}{\rotatebox{90}{Uranus}}
& Rings     & $\thicksim$500 & 500 & 0.68897 & 0.47873 & -0.54651 \\ %& 492.9 & 0.20924 & 1.30798 & -1.20411 & 1.21505 & -0.67196 & 0.14381 \\
& Portia \textit{group} & $\thicksim$500 & 500 & 0.68897 & 0.47873 & -0.54651 \\ %& 492.9 & 0.20924 & 1.30798 & -1.20411 & 1.21505 & -0.67196 & 0.14381 \\
& Ariel     & $\thicksim$600 & 600 & 0.78074 & 0.38427 & -0.53777 \\ %& 610.9 & 0.30854 & 1.36620 & -1.83572 & 2.33221 & -1.63082 & 0.45959 \\
& Titania   & $\thicksim$600 & 600 & 0.78074 & 0.38427 & -0.53777 \\ %& 610.9 & 0.30854 & 1.36620 & -1.83572 & 2.33221 & -1.63082 & 0.45959 \\
& Oberon    & $\thicksim$600 & 600 & 0.78074 & 0.38427 & -0.53777 \\ %& 610.9 & 0.30854 & 1.36620 & -1.83572 & 2.33221 & -1.63082 & 0.45959 \\
\hline
\multirow{4}{1em}{\rotatebox{90}{Neptune}}
& Fraternit\'e & $\thicksim$500 & 500 & 0.68897 & 0.47873 & -0.54651 \\ %& 492.9 & 0.20924 & 1.30798 & -1.20411 & 1.21505 & -0.67196 & 0.14381 \\
& Egalit\'e   & $\thicksim$500 & 500 & 0.68897 & 0.47873 & -0.54651 \\ %& 492.9 & 0.20924 & 1.30798 & -1.20411 & 1.21505 & -0.67196 & 0.14381 \\
& Nereid    & $\thicksim$570 & 560 & 0.75079 & 0.41593 & -0.54334 \\ %& 559.9 & 0.26892 & 1.34319 & -1.58427 & 1.91271 & -1.31530 & 0.37295 \\
& Triton    & $\thicksim$400 & 400 & 0.16732 & 0.84347 & -0.03516 \\ %& 401.9 & 0.12323 & 1.08648 & -0.43974 & 0.45912 & -0.32759 & 0.09850 \\
\hline     
\end{tabular}}%}
\end{center}
\end{table}

%%%%%%%%%%%%%%% END OF DOCUMENT %%%%%%%%%%%%%%%%%%%%%%%%%
\end{document}